\documentclass[journal=nalefd,manuscript=letter]{achemso}

\usepackage{xcolor}
\definecolor{green2}{rgb}{0.0, 0.5, 0.0}

\usepackage[normalem]{ulem}

\usepackage[version=3]{mhchem} 
\usepackage{achemso}
\mciteErrorOnUnknownfalse
\usepackage[hidelinks]{hyperref}
\usepackage{xcolor}
\hypersetup{
    colorlinks,
    linkcolor={blue},
    citecolor={blue},
    urlcolor={blue}
}
\DeclareUnicodeCharacter{03C9}{$\omega$}
\DeclareUnicodeCharacter{2212}{-}
\usepackage{bm}
\usepackage{orcidlink}

\newif\ifptitle
\newif\ifpnumber
\newcounter{para}

\author{Lucy S. Nathwani\,\orcidlink{0009-0001-5002-3812}}
\affiliation{Department of Physics, Harvard University, Cambridge, MA, 02138}

\author{Anne Ruperto\,\orcidlink{0009-0002-9271-7131}}
\affiliation{Department of Physics, Harvard University, Cambridge, MA, 02138}
\author{Ashvini Vallipuram\,\orcidlink{0009-0003-8895-4289}}
\affiliation{Department of Physics, Harvard University, Cambridge, MA, 02138}
\author{Abigail Y. Jiang\,\orcidlink{0000-0001-8003-8380}}
\affiliation{John A. Paulson School of Engineering and Applied Sciences, Harvard University, Cambridge, MA, 02138}
\alsoaffiliation{Department of Physics, Harvard University, Cambridge, MA, 02138}
\author{Grace A. Pan\,\orcidlink{0000-0002-4512-1215}}
\affiliation{Department of Physics, Harvard University, Cambridge, MA, 02138}
\author{Dan Ferenc Segedin\,\orcidlink{0000-0001-7162-8100}}
\affiliation{Department of Physics, Harvard University, Cambridge, MA, 02138}
\author{Ari B. Turkiewicz\,\orcidlink{0000-0001-5729-0289}}
\affiliation{Department of Physics, Harvard University, Cambridge, MA, 02138}
\author{Charles M. Brooks\,\orcidlink{0000-0001-9087-7321}}
\affiliation{Department of Physics, Harvard University, Cambridge, MA, 02138}
\author{Jarad A. Mason\,\orcidlink{0000-0003-0328-7775}}
\affiliation{Department of Chemistry and Chemical Biology, Harvard University, Cambridge, MA, 02138}
\email{mason@chemistry.harvard.edu}
\author{Qichen Song\,\orcidlink{0000-0002-1090-4068}}
\affiliation{Department of Chemistry and Chemical Biology, Harvard University, Cambridge, MA, 02138}
\email{qichensong@g.harvard.edu}
\author{Julia A. Mundy\,\orcidlink{0000-0001-8454-0124}}
\affiliation{Department of Physics, Harvard University, Cambridge, MA, 02138}
\alsoaffiliation{John A. Paulson School of Engineering and Applied Sciences, Harvard University, Cambridge, MA, 02138}
\email{mundy@fas.harvard.edu}

\title[FDTR]
  {Observation of microscopic domain effects in the metal-insulator transition of thin-film NdNiO$_3$}

\abbreviations{IR,NMR,UV}

\begin{document}


\begin{tocentry}

    \includegraphics[width=7.5cm]{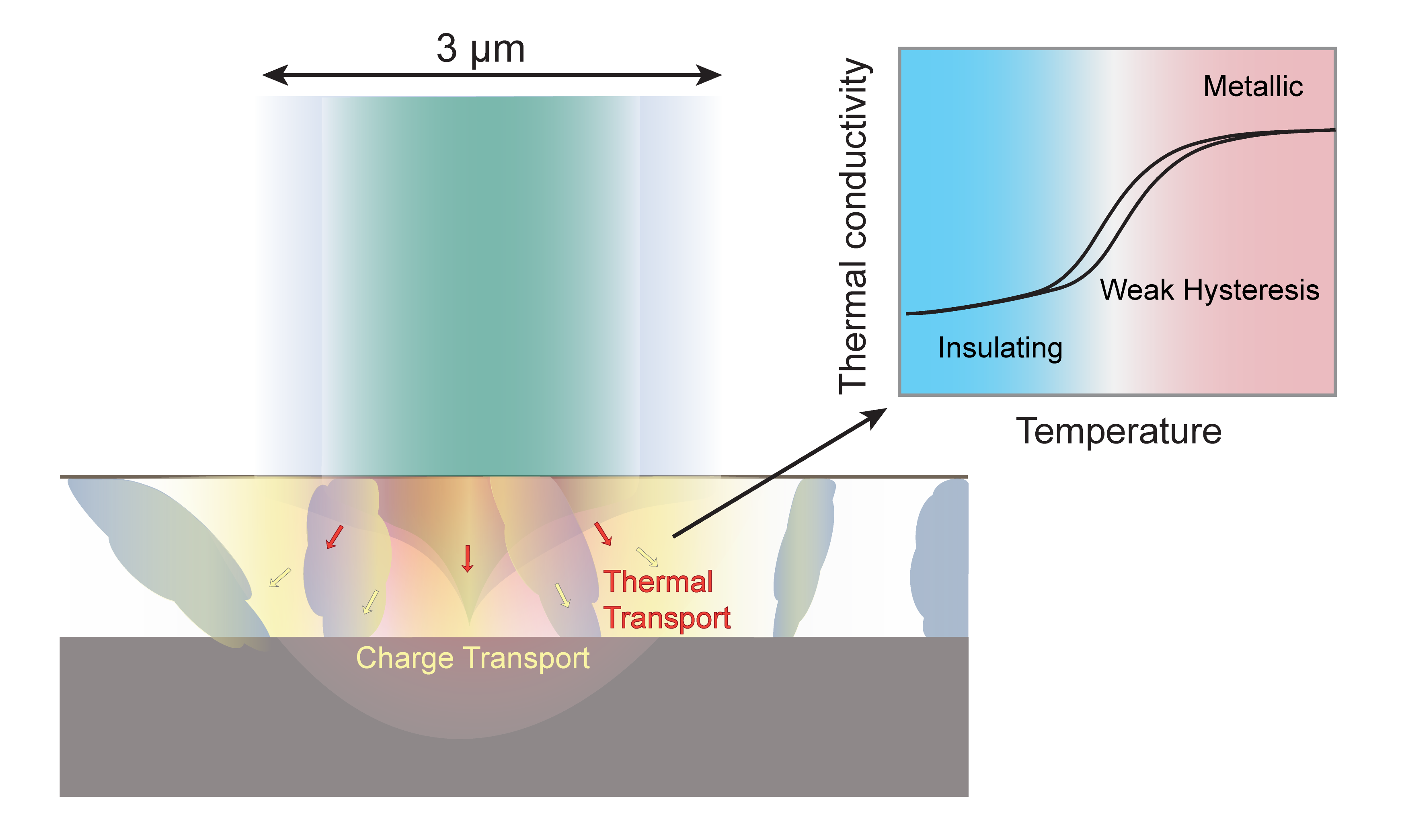}    
    \label{fig:toc}
\end{tocentry}

\begin{abstract} 
Perovskite oxides display correlated electrical, magnetic, and thermal properties that can be further tuned in the thin-film limit, making them contenders for next-generation electronics. Measuring thermal transport in thin films is challenging, because traditional techniques are dominated by the substrate. Here, frequency-domain thermoreflectance (FDTR) of an epitaxial NdNiO$_3$ thin film reveals a sharp change in out-of-plane thermal conductivity across the metal–insulator transition. Complementary frequency-domain photoreflectance (FDPR) reveals a large change in ambipolar diffusivity of photoexcited carriers. While the in-plane electrical resistance shows large hysteresis, out-of-plane thermal and charge transport shows negligible hysteresis. We attribute this discrepancy to anisotropy in the percolation of nanoscale domains across the transition as the film thickness approaches the domain length scale. We establish FDTR and FDPR as sensitive probes of quantum material phase transitions and highlight NdNiO$_3$ for thermal control and memory applications.

\end{abstract}

As the demand for miniaturized, energy-efficient electronics grows, materials that undergo temperature-driven metal–insulator transitions (MIT) have emerged as candidates for next-generation technologies, including thermal switches for heat regulation, smart windows for thermochromic glazing, nonvolatile memristors for data storage, and neuromorphic computing elements for brain-inspired computing\cite{wen_review_2024, haddad_review_2022,xia_memristive_2019, imada_metal-insulator_1998}. Among these, rare-earth perovskite nickelates, \textit{R}NiO$_3$ (\textit{R} = La, Pr, Nd...), are model systems due to their highly tunable phase transitions \cite{catalano_rare-earth_2018,alonso_room-temperature_2000,garcia-munoz_structure_2009,liu_heterointerface_2013}. In particular, neodymium nickelate, NdNiO$_3$, undergoes a first-order phase transition from a paramagnetic metallic state to an antiferromagnetic insulating state, with a concomitant change in the crystal structure at the highest temperature of the \textit{R}NiO$_3$ series. It exhibits thermal hysteresis due to percolation across domains that emerge during the transition \cite{preziosi_direct_2018,mattoni_striped_2016,lee_imaging_2019}. 
In single crystal NdNiO$_3$, this transition occurs near 200 K. However, in the thin-film limit, the transition temperature can be tuned through epitaxial strain and film thickness \cite{zhang_key_2016, laffez_evidence_2004, catalano_rare-earth_2018, wang_oxygen_2016, shi_thermodynamics_2024}, making it highly adaptable for integration into electronic devices. While most studies have focused on the electrical and magnetic properties of thin film NdNiO$_3$, its thermal properties are highly relevant for its practical applications in devices. Fully exploiting these materials in next-generation devices, however, requires an understanding of anisotropic charge and heat transport at the nanoscale, particularly as the thickness of the film approaches the length scales of domain formation during the phase transition.

Measuring thermal transport in ultrathin films remains challenging due to the limited sensitivity of conventional bulk techniques, as traditional measurements of thermal conductivity using an applied heat source and a voltage measurement are dominated by substrate contributions \cite{grissonnanche_electronic_2024}. While $3\omega$ techniques offer improved sensitivity to thin films, they require complex and invasive sample preparation and lack spatial resolution \cite{cahill_thermal_1990}. Noncontact optical reflectance-based techniques provide an alternative avenue that addresses these limitations \cite{oh_thermal_2010}. In particular, frequency-domain thermoreflectance (FDTR)\cite{yang_thermal_2013, schmidt_frequency-domain_2009} is capable of measuring the thermal conductivity of films as thin as several tens of nanometers, while requiring only minimal sample preparation (i.e., depositing a metallic transducer). FDTR can even measure thermal transport in monolayer materials like graphene, depending on the film composition and the substrate \cite{yang_imaging_2015,akura_frequency-domain_2025}, and it can be used to map thermal properties with micrometer resolution (Figure~S4). The same optical setup without a transducer layer can be used to perform frequency-domain photoreflectance (FDPR), further allowing the charge and thermal transport to be probed simultaneously \cite{song_probing_2024}. Additionally, FDTR is particularly sensitive to the cross-plane thermal conductivity. Scanning thermal microscopy (SThM) techniques have been used to measure the cross-plane thermal conductivity at higher spatial resolution, but these techniques require specialized sample preparation and calibration to gain sensitivity to multiple sample thicknesses simultaneously \cite{gonzalezmunoz_direct_2023}. With FDTR and FDPR, we can obtain ambipolar diffusivity with no sample preparation and thermal conductivity by depositing only a transducer layer.

In this work, we use FDTR to measure the thermal conductivity of a 57.5 nm thick epitaxial NdNiO$_3$ film grown on a LaAlO$_3$ substrate across its metal–insulator transition, the thinnest such epitaxial oxide measured with FDTR to date. While we observe strong hysteresis in the in-plane electrical transport, consistent with prior measurements \cite{preziosi_direct_2018,zhang_key_2016,hooda_electronic_2016}, we observe significantly reduced hysteresis in the thermal conductivity during heating and cooling cycles. We also perform the first demonstration of FDPR measurements on a thin film over a wide modulation-frequency bandwidth; we model the laser-induced charge transport as ambipolar diffusion in which electrons and holes diffuse according to their density gradients before recombination. We observe a significant change in the ambipolar diffusivity across the transition, accompanied by weaker hysteresis. The suppressed hysteresis in both thermal and charge transport is attributed to geometric anisotropy in the thin-film architecture, where the domain size is constrained by the film thickness, limiting percolation in the out-of-plane direction. These findings demonstrate the power of a local probe to resolve the nanoscale phase separation that governs the dynamics of tunable MITs in thin-film correlated oxides.

\begin{figure}[h]
    \centering
    \includegraphics[width=\linewidth]{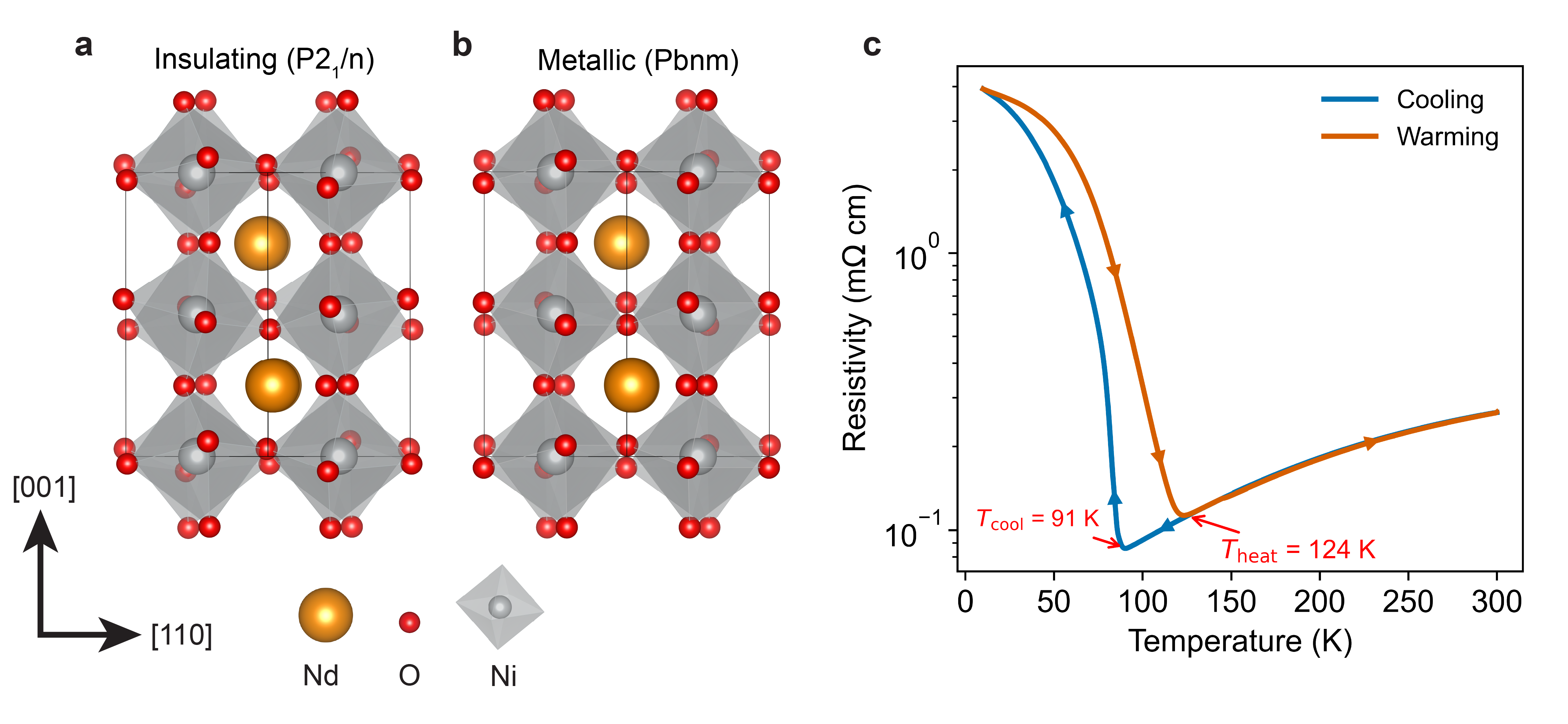}
    \caption{
    (a) Distorted perovskite structure of NdNiO$_3$ in the insulating phase. (b) The structure in the metallic phase. The changes in octahedral tilts drive the transition\cite{garcia-munoz_structure_2009}. (c) Temperature-dependent resistance capturing the MIT (see SI). $T_\mathrm{MIT}$ is 91 K during cooling and 124 K during heating. The film is 57.5 nm (150 unit cells) thick in the (001)$_\mathrm{PC}$ direction, grown on LaAlO$_3$ (100)$_\mathrm{PC}$.}
    \label{fig:nickelate}
\end{figure}

In bulk single crystals of NdNiO$_3$, the MIT is accompanied by a subtle structural transition from orthorhombic ($\mathrm{Pbnm}$) to monoclinic ($\mathrm{P2_1/n}$) symmetry (Figure~\ref{fig:nickelate}a,b)\cite{garcia-munoz_structure_2009,catalano_rare-earth_2018,Gomes_structural_2024}; the transition has a thermal hysteresis of $\sim$4 K. In epitaxial thin films, the hysteresis increases to $\sim$30 K due to epitaxial strain \cite{catalan_metal-insulator_2000,lee_imaging_2019}. We synthesize thin-film NdNiO$_3$ using ozone-assisted molecular-beam epitaxy (MBE) for atomic-layer control of the film deposition (see Figures~S1-S3). Electrical transport measurements of our thin-film sample show a sharp change from metallic to insulating behavior in bulk resistivity measurements upon cooling at $T_{\mathrm{cool}}$ = 91 K, and we observe the reverse change at $T_{\mathrm{heat}}$ = 124 K during heating (Figure~\ref{fig:nickelate}c). The biaxial compressive strain imposed by the LaAlO$_3$ substrate lowers $T_{\text{MIT}}$ by over 70 K, consistent with prior observations of strain-induced suppression of $T_{\text{MIT}}$ \cite{zhang_key_2016,catalano_rare-earth_2018}. 

\begin{figure}[t!]
    \centering
\includegraphics[width=\linewidth]{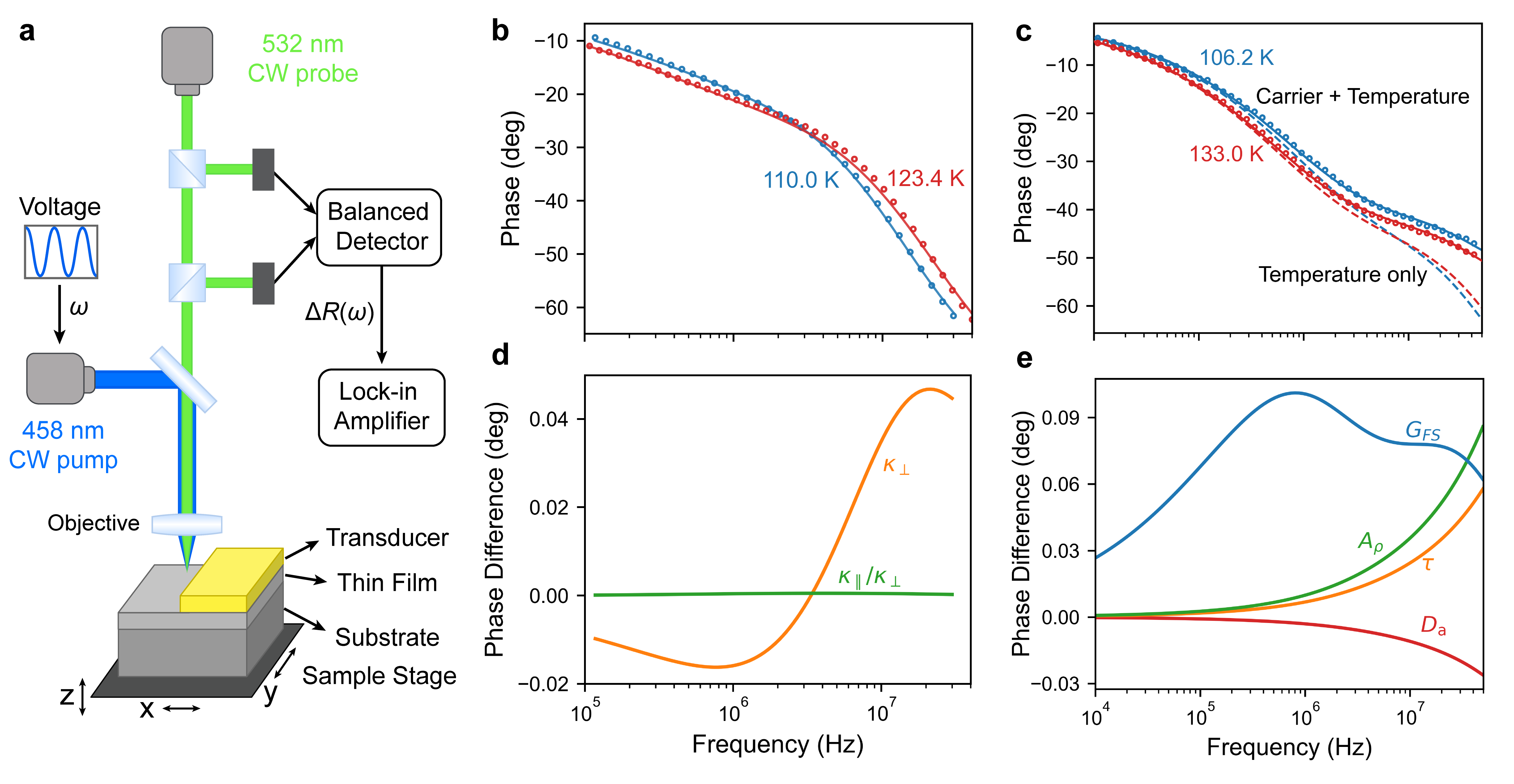}
    \caption{ 
    (a) Schematic of an integrated frequency-domain thermoreflectance (FDTR) and frequency-domain photoreflectance (FDPR) measurement setup. A continuous wave (CW) pump laser at 458 nm, power-modulated at a frequency $\omega/2\pi$, excites the sample. A CW probe laser at 532 nm reads out the perturbed reflectance. Both laser beams are focused to a 3 \textmu m beam waist radius. The reference and reflected beams are first balanced and then detected by a balanced detector. The change in reflectance ${\Delta R(\omega)}$ is extracted from the output signal by a lock-in amplifier. FDTR and FDPR measurements are performed at the gold-coated (65.7 nm) and bare regions of the sample, respectively, with a piezoelectric stage being used to translate the sample in the xy-plane and to locate the beam focus along the z-axis. (b) FDTR phase data (circles) in the metallic (123.4 K) and insulating (110.0 K) phases, along with best fits using the Fourier heat conduction model (lines). (c) FDPR data in the metallic (133.0 K) and insulating (106.2 K) phases, along with best fits using the carrier and temperature model (solid lines); the phases of the temperature part of the signal are plotted in dashed lines. (d, e) Sensitivity analysis for FDTR (d) and FDPR (e) data fitting at 110.0 and 133.3 K, respectively. Specifically, $\kappa_\parallel$ and $\kappa_\perp$ correspond to the in-plane and out-of-plane thermal conductivity, respectively. $G_{FS}$ is the thermal boundary conductance between the NdNiO$_3$ film and the LaAlO$_3$. $A_\rho$ is related to the magnitude of the carrier contribution to the signal, $\tau$ is the carrier recombination time, and $D_\mathrm{a}$ is the ambipolar diffusivity. The phase difference is defined as the change in phase resulting from a + 1\% relative increase in the corresponding variable.}
    \label{fig:setup}
\end{figure}

We explore thermal and electronic transport through complementary techniques of FDTR and FDPR on the same sample. As illustrated in Figure~\ref{fig:setup}a, to isolate the thermal response of the film, we perform FDTR measurements for a region coated with a gold transducer layer (Figure~S2,~S3), which immediately converts electronic excitations into heat and ensures that the measured change in reflectance $\Delta R(\omega)$ is based only on the thermal properties of the sample (see SI)\cite{yang_thermal_2013, schmidt_frequency-domain_2009}. To measure charge transport, we perform FDPR measurements on an adjacent uncoated region of the same sample; the photoreflectance signal includes contributions from heat diffusion and charge transport. We then determine the ambipolar carrier diffusivity across the transition.

In FDTR, the thermoreflectance signal is proportional to the change in the surface temperature of the gold transducer. The relative phase between the oscillations of the thermoreflectance signal and the pump modulation exhibits a distinct frequency dependence in the metallic phase compared to the insulating phase (Figure~\ref{fig:setup}b). According to the sensitivity analysis based on the Fourier heat conduction model, our FDTR measurement is exclusively sensitive to the out-of-plane thermal conductivity $\kappa_\perp$ (Figure~\ref{fig:setup}d). At high modulation frequencies, the phase shifts toward zero with increasing $\kappa_\perp$, while at low modulation frequencies, it shifts to more negative values with increasing $\kappa_\perp$ (Figure~\ref{fig:setup}b). This relationship suggests that the out-of-plane thermal conductivity of the metallic phase is higher than that of the insulating phase, as expected.

In FDPR, the photoreflectance from the bare sample is the sum of the carrier-induced and temperature-induced components, both of which are represented by complex numbers. As shown in Figure~\ref{fig:setup}c, at higher modulation frequencies, the carrier contribution can lower the overall phase lag compared with the temperature contribution alone, as carrier diffusion typically occurs much faster than heat diffusion. The different phase curves in the metallic and insulating states, especially at high modulation frequencies, are closely related to the difference in the carrier part of the signal. From the sensitivity analysis in Figure~\ref{fig:setup}e, we observe that the constant $A_\rho$ that determines the magnitude of the carrier contribution to the signal, the carrier recombination time $\tau$, and ambipolar diffusivity $D_\mathrm{a}$ are correlated, allowing us to fit only one of these three parameters reliably. Given that $D_\mathrm{a}$ is substantially more susceptible to changes in transport than both $A_\rho$ and $\tau$, we fix the values of $A_\rho$ and $\tau$ and treat $D_\mathrm{a}$ along with the thermal boundary conduction between the thin film and substrate, $G_{FS}$, as fitting variables (see Figure~S7,~S8).

\begin{figure}[t]
    \centering
    \includegraphics[width=\linewidth]{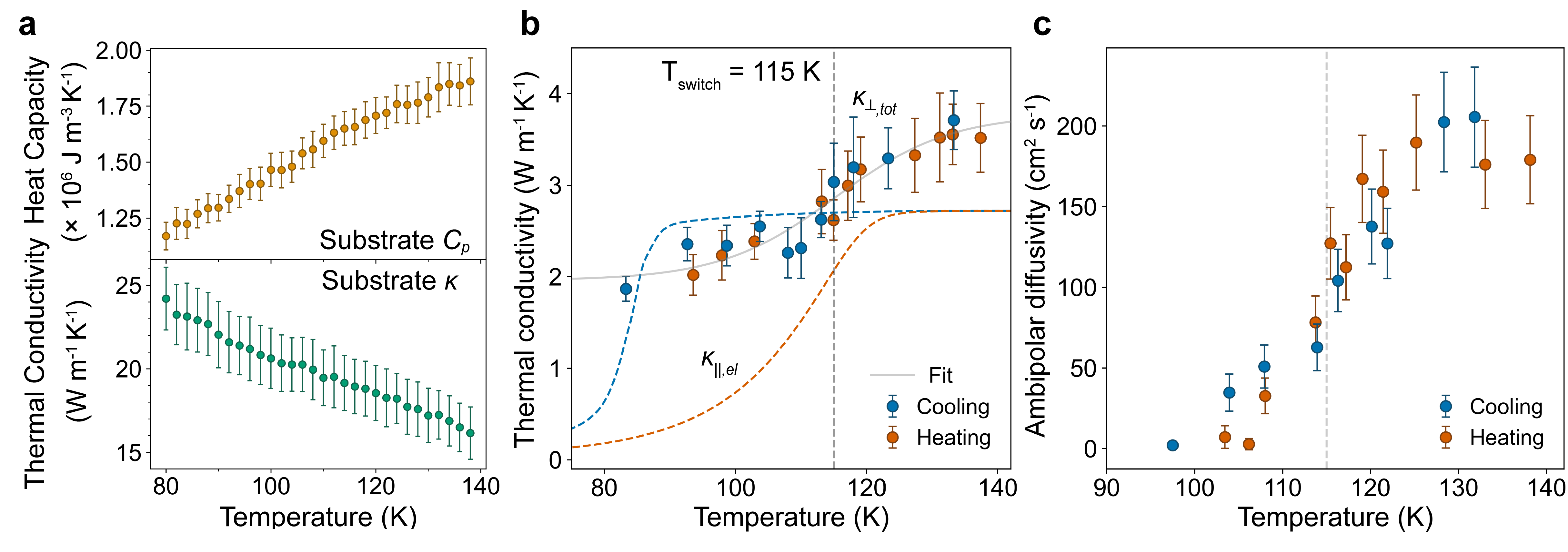}
    \caption{
    (a) The thermal conductivity and heat capacity of the substrate LaAlO$_3$ measured using FDTR. (b) Out-of-plane thermal conductivity, $\kappa_{\perp,\text{tot}}$, of thin film NdNiO$_3$ as a function of temperature, measured using FDTR over three cooling and two heating cycles (see~Figure~S5 for the full data and Table~S1 for numerical values). The electronic contribution to in-plane thermal conductivity, $\kappa_{\parallel,\text{el}}$ predicted using the Wiedemann–Franz law is plotted as dashed lines. The hysteresis is weaker in the out-of-plane direction than in the in-plane direction, highlighting the anisotropy of the thin-film geometry. A modified logistic function, $f(T)=a\left[1+e^{-b\left(T-T_\mathrm{switch}\right)}\right]^{-1}+c$, is used to fit the thermal conductivity trend (gray line). The thermal switching temperature Tswitch is found to be 115 K. (c) The ambipolar diffusivity of NdNiO$_3$, measured using FDPR, across the metal–insulator transitions during cooling and heating. The dashed line corresponds to $T_\mathrm{switch}=$ 115 $\pm$ 4 K, with the standard deviation obtained from a bootstrap analysis (Figure~S6). Fixed parameters $A_\rho$ = $8\times10^{-25}$ m$^3\,$K and $\tau$ = 0.11 ns are chosen to minimize fitting residuals across all temperatures.} 
    \label{fig:kappa_d}
\end{figure}

Analyzing FDTR and FDPR data to extract transport properties of the thin film requires knowledge of the substrate properties. Therefore, we characterized the thermal properties of the LaAlO$_3$ substrate independently. We performed FDTR measurements on a LaAlO$_3$ substrate coated with a transducer layer (Figure~S14,~S15) and measured its thermal conductivity and heat capacity over the temperature range of 80–140 K (Figure~\ref{fig:kappa_d}a). The measured substrate heat capacity and thermal conductivity (Figure~\ref{fig:kappa_d}a) agree well with literature values for the heat capacity and thermal conductivity of LaAlO$_3$\cite{schnelle_specific_2001}. Notably, our approach enables the simultaneous determination of both thermal conductivity and heat capacity from a single measurement (Figure~S13). This is achieved by varying the modulation frequency over a wide range: at higher frequencies, the thermoreflectance’s phase has a positive correlation with both thermal conductivity and heat capacity, while at lower frequencies, its dependences on thermal conductivity and heat capacity have opposite signs.

We then fix the substrate thermal conductivity and heat capacity in our heat conduction model and determine the out-of-plane thermal conductivity ($\kappa_{\perp,\text{tot}}$) of the NdNiO$_3$ thin film. We perform heating and cooling measurements between 80–140 K. Simultaneously fitting both $\kappa$ and $C_p$ for the film is challenging because of the high uncertainty and because these parameters are strongly correlated. Therefore, we fix the film’s heat capacity using a Debye model derived from bulk heat capacity measurements on NdNiO$_3$ (see Figure~S16)\cite{hooda_electronic_2016,barbeta_metal-insulator_2011}. With only the film’s thermal conductivity as a free parameter, we achieve stable fits with uncertainty $<18$\%. Figure~\ref{fig:kappa_d}b shows the thermal conductivity $\kappa$ of the thin film NdNiO$_3$ during cooling and heating cycles. 

During cooling, the thermal conductivity $\kappa_{\perp,\text{tot}}$ of NdNiO$_3$ gradually decreases between 133.3 and 123.3 K. Between 123.3 and 110.0 K, it decreases more sharply by 33\%, a phenomenon we describe as thermal switching. We compare the out-of-plane thermal conductivity to the predicted electronic contribution to the in-plane thermal conductivity using the Wiedemann–Franz law. It is important to recognize that the Wiedemann–Franz law can fail in correlated oxides \cite{PhysRevLett.97.067005,lee_anomalously_2017}, where non-Fermi-liquid transport might emerge \cite{PhysRevB.88.125107}. Consequently, we regard the predicted electronic thermal conductivity as a comparative indicator of in-plane heat transport. We calculate $\kappa_{\parallel,\text{el}}=L\sigma T$, where $L$ is the Lorenz number, $\sigma$ is the separately measured in-plane electrical conductivity (Figure~\ref{fig:nickelate}a) and $T$ is the temperature. In principle, we expect no anisotropy in the properties of bulk NdNiO$_3$, so we attribute the differences in thermal conductivity to finite scaling in the thin film limit.

On cooling, we observe the onset of thermal switching in $\kappa_{\perp,\text{tot}}$ more than 20 K higher in temperature than in $\kappa_{||,\text{el}}$. The heating curve of $\kappa_{\perp,\text{tot}}$ closely matches the cooling curve across the transition, suggesting that this transition temperature anisotropy arises from weak hysteresis in the out-of-plane direction. The hysteresis remains weak across different pump laser powers, eliminating the possibility that the laser reduces the hysteresis (Figure~S9), and we can also rule out a photoinduced transition (Figure~S11). Consistent with this behavior, the out-of-plane ambipolar diffusivity during heating and cooling from FDPR decreases sharply from 120 to 110 K and changes by much smaller amounts for $T$ $<$ 110 K (Figure~\ref{fig:kappa_d}c), which reinforces the idea that an electronic phase transition occurs at $T_\mathrm{switch}$. The DC contribution of the reflectance from the bare sample, related to the dielectric constants of the thin film, also shows a sharp discontinuity at around 115 K, providing additional evidence for the electronic phase transition occurring near $T_{\mathrm{switch}}$ (Figure~S12). Together, these measurements show that the transition temperature and the associated hysteresis measured in the out-of-plane direction differ substantially from those observed in the in-plane direction. 

 Both epitaxial strain and film thickness can tune the MIT in NdNiO$_3$, so we first consider whether differing strains in the out-of-plane direction can lead to anisotropic hysteresis. We expect that the film is under compressive strain in the in-plane direction due to the substrate, and therefore, it is under tensile strain in the out-of-plane direction. \cite{wang_competition_2015}. While bulk NdNiO$_3$ has a hysteresis loop width of $\sim$4 K\cite{catalan_metal-insulator_2000}, both compressive and tensile strain on thin films increase the hysteresis loop width to $>$ 10 K\cite{scherwitzl_electricfield_2010}. We therefore expect hysteresis in the in-plane and out-of-plane directions, and anisotropic strain effects should not eliminate hysteresis in the out-of-plane direction. To understand the origin of the discrepancy, we then consider the length scale of the FDTR and FDPR measurements compared to electrical resistivity measurements, understanding the dimensionality of the transition in the thin-film limit.

Previous imaging and local transport studies of the NdNiO$_3$ metal-insulator transition (MIT) show that the transition proceeds through mesoscale phase coexistence with insulating and metallic domains whose lateral size and morphology evolve strongly with temperature. Epitaxial strain has a strong influence on both the transition temperature and hysteresis width, so quantitative transition temperatures reported in the literature vary significantly between samples. In some epitaxial films, insulating domains form stripe-like patterns aligned with the substrate terracing, while in others such striping is not apparent. In the striped case, insulating domains are typically $\sim$200 nm wide and can extend much longer along the terrace direction \cite{mattoni_striped_2016}, whereas in nonstriped films the domains appear more irregular but retain a comparable characteristic lateral length scale \cite{bisht_phase_2017,preziosi_direct_2018}. Upon cooling through the MIT, isolated insulating domains nucleate at temperatures slightly above the nominal transition. As temperature decreases further, these insulating regions expand and connect with neighboring domains, progressively reducing the connectivity of the metallic network \cite{preziosi_direct_2018,bisht_phase_2017}. This loss of metallic connectivity drives the sharp increase in the resistivity observed on cooling. In films that exhibit stripe ordering, the insulating domain coverage increases with decreasing temperature and saturates at roughly 60\%, while the striped morphology remains visible; metallic domains persist, but their lateral connectivity is strongly suppressed \cite{mattoni_striped_2016}. Upon heating, the evolution is different: insulating domains tend to melt from their edges and pinch off into smaller, disconnected regions, producing a domain distribution that is qualitatively distinct from that observed during cooling, even at the same insulating fraction. Because electrical transport depends on domain connectivity rather than phase fraction alone, this asymmetry in domain morphology and connectivity gives rise to observed macroscopic transport hysteresis.

\begin{figure}[t!]
    \centering
    \includegraphics[width=0.9\linewidth]{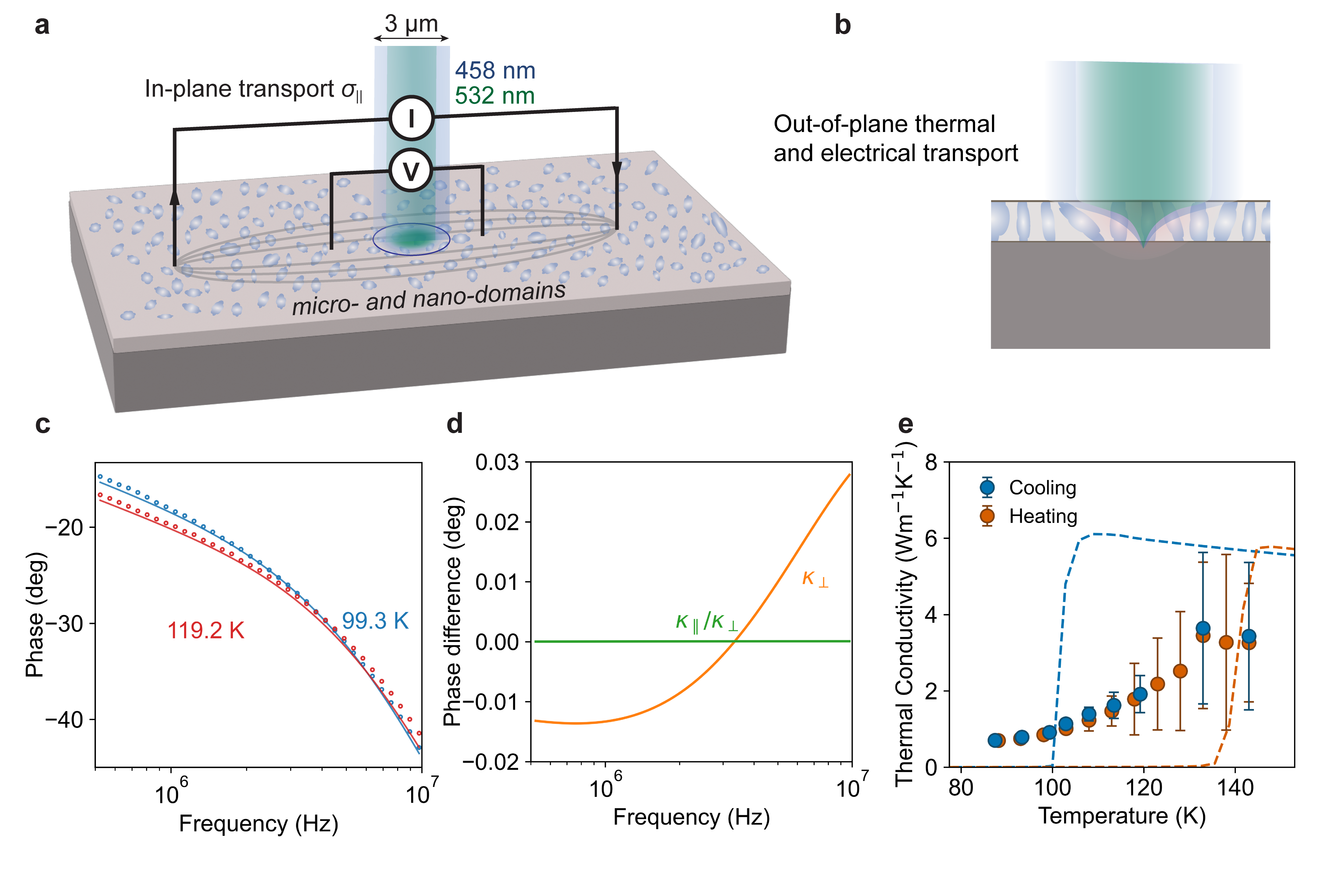}
    \caption{(a) Qualitative (not to scale) schematic representing the formation of insulating domains in NdNiO$_3$ near the phase transition temperature. FDPR probes out-of-plane electron transport on a length scale covering several domains, whereas the electrical resistance measurement aggregates in-plane contributions from electrons percolating across on the order of 10$^3$ domains. (b) Cross-sectional view of the beam penetration and domain formation (not to scale). Domains form instantaneously across the film thickness, directly creating single-phase transport channels without metal–insulator interfaces; hence, out-of-plane thermal and electron transport exhibits low hysteresis. (c) Phase lag data as a function of modulation frequency for the metallic (red) and insulating (blue) phases, together with best-fit model curves, for the 20.6 nm sample. (d) Sensitivity analysis for phase lag best fits at 133 K. A 1\% increase in thermal conductivity produces negligible changes for the in-plane component. As with the 57.5 nm sample, the sensitivity to the in-plane thermal conductivity is nearly zero, while the sensitivity to the out-of-plane thermal conductivity is reduced. This reduction arises from the decreased film thickness and the correspondingly lower thermal resistance. (e) Out-of-plane thermal conductivity as a function of temperature, extracted from best-fit phase lag curves. Wiedemann–Franz law predictions for the electron contribution to in-plane thermal conductivity are overlaid (dashed lines). Minimal hysteresis is again observed in the out-of-plane thermal conductivity between the cooling and heating measurements, consistent with the 57.5 nm film.} 
    \label{fig:domains}
\end{figure}

The exact domain morphology in our films across the transition is not known, but we can estimate the length scale at which domains affect thermal and electrical transport. During heating, the effective metallic transport domains in the coexistence regime are typically 100–300 nm in size, as inferred from conductive atomic force microscopy (AFM) and related measurements \cite{preziosi_direct_2018}. Consistent with this scale, nanogap transport experiments report pronounced deviations from bulk hysteresis behavior for gap sizes in the range of $\sim$40–150 nm, where resistance drops abruptly when a single metallic domain bridges the contacts \cite{lee_imaging_2019}. These characteristic length scales are comparable to our film thickness (57.5 nm), suggesting that out-of-plane transport in thin films is governed by the formation of vertically continuous metallic paths rather than by in-plane percolation (Figure~\ref{fig:domains}b). Consequently, out-of-plane transport is not as susceptible to domain connectivity, which changes the hysteresis. The FDTR/FDPR measurements probe reflectance over a spot size of $\sim$3\textmu m. At these length scales, the measurement aggregates over hundreds of lateral domains within the coexistence regime (Figure~\ref{fig:domains}a). As a result, abrupt resistivity or conductivity changes associated with the nucleation or percolation of individual metallic domains in the out-of-plane direction are averaged out, creating a smooth change in thermal and electrical transport without hysteresis.

This phenomenon is further corroborated by another measurement on a 20.6 nm thick NdNiO$_3$ film on LaAlO$_3$, also synthesized with MBE (Figure~S10). This thinner film demonstrates a similar lack of hysteresis in the out-of-plane thermal conductivity. Figure~\ref{fig:domains}c shows phase lag data and the best fits for metallic and insulating states. The sensitivity analysis (Figure~\ref{fig:domains}d) shows that FDTR is less sensitive to this sample’s out-of-plane thermal conductivity, $\kappa_\perp$, than to that of the 57.5 nm sample, as the 20.6 nm film has a lower thermal resistance. Still, we observe no hysteresis in the out-of-plane thermal conductivity measurements(Figure~\ref{fig:domains}e). This result matches that of the 57.5 nm film, again suggesting a lack of percolation in the out-of-plane direction. We can therefore conclude that the anisotropy in the hysteresis is common to NdNiO$_3$ films of similar thicknesses to the 57.5 nm sample.

FDTR and FDPR in a thin-film geometry are uniquely sensitive to domain emergence in a phase transition, capturing the early onset of localized phase separation that precedes changes in the macroscopic properties. Although the diffraction limit prevents us from directly resolving the thermal conductivity of a single domain or right at the interface, our approach uniquely provides access to the cross-plane thermal conductivity, which, when combined with the mapped domain structures reported in previous studies \cite{bisht_phase_2017,Gomes_structural_2024,li_scale-invariant_2019,mattoni_striped_2016}, allows us to understand how domains govern transport properties.

Our study reveals a microscopic, domain-mediated phase transition in NdNiO$_3$ preceding the MIT transition observed via bulk transport measurement. Thermal switching occurs at a temperature significantly above the metal–insulator transition identified by in-plane electrical transport measurements, with significantly reduced hysteresis. The thin-film geometry plays a significant role in the reduced hysteresis as the length scale of the domains formed during the transition is larger than the film thickness, allowing us to reduce percolation across domains in the out-of-plane direction. While much research on NdNiO$_3$ has been dedicated to tuning the hysteresis and transition temperature of the MIT, the anisotropy of the thin-film geometry has yet to be addressed and presents another way to tune the properties of the MIT. These results position NdNiO$_3$ thin films within a growing class of correlated oxides that exhibit nontrivial transport behavior across electronic phase transitions. Violations of the Wiedemann–Franz law observed in VO$_2$ nanowires near the MIT \cite{lee_anomalously_2017} suggest the involvement of unconventional quasiparticles or other strongly correlated transport phenomena \cite{zhang_anomalous_2017}. The unusual behavior of the reduced dimensionality in VO$_2$ nanowires highlights the macroscopic effects of fundamental length scales in correlated oxides with phase transitions. Our findings open new avenues for exploring the role of geometrical confinement and domain structure in such materials, advancing NdNiO$_3$ thin films as compelling candidates for realizing passive thermal switches and nonvolatile memristors.

Our work demonstrates that FDTR and FDPR offer unique capabilities for probing phase transitions in thin films where the domain sizes exceed the film thickness. In contrast to bulk resistivity, which probes in-plane transport percolating across many domains, our reflectance-based measurements are more sensitive to the local emergence and distribution of metallic and insulating domains. FDTR measurements in ultrathin films, therefore, primarily average transport across single domains, minimizing percolative effects and hysteresis. The sharp features observed in both thermal transport and ambipolar diffusivity near $T_\mathrm{switch}$ with weak hysteresis confirm the differences in the measurement due to the geometry of FDTR and FDPR. These insights demonstrate that FDTR is a powerful probe of phase transitions in complex oxides, enabling access to out-of-plane transport properties in regimes inaccessible to traditional bulk methods.

Beyond this central result, we introduce several methodological advances. We demonstrate that FDTR can simultaneously measure the thermal conductivity and heat capacity of a bulk substrate material (in this case, LaAlO$_3$) across a wide temperature range. We also report the first use of FDPR to extract ambipolar carrier diffusivity in a thin film system. The observed drop in $D_\mathrm{a}$ across the same temperature range as the thermal transition confirms the coevolution of electronic and thermal transport across the MIT. This combined FDTR/FDPR framework opens avenues for studying nanoscale phase transitions, domain dynamics, and anisotropic transport in complex oxides, both in equilibrium and under external stimuli such as strain or electric fields.

\begin{acknowledgement}
The authors thank Aaron J. Schmidt for helpful discussions. L.S.N. thanks Suzanne Smith and Jennifer Hoffman for their feedback on the manuscript. This project was primarily supported by the US Department of Energy, Office of Basic Energy Sciences, Division of Materials Sciences and Engineering, under Award No. DE-SC0021925. We also acknowledge support from the National Science Foundation with award DMR-2323970. 
This work was performed in part at the Harvard University Center for Nanoscale Systems (CNS); a member of the National Nanotechnology Coordinated Infrastructure Network (NNCI), which is supported by the National Science Foundation under NSF award no. ECCS-2025158. Q.S., L.S.N., and A.R. acknowledge support from the Harvard Quantum Initiative. 
A.Y.J., G.A.P., and D.F.S. acknowledge support from the NSF Graduate Research Fellowship Grant DGE-1745303. A.Y.J. and G.A.P. were also supported by the Paul \& Daisy Soros Fellowship for New Americans, and A.Y.J. by the Ford Foundation. J.A.M. acknowledges funding from the Star Friedman Fund at Harvard University. L.S.N. and A.R. acknowledge support from the Herchel Smith Undergraduate Science Research Program. Generative A.I. was used in the code to create plots for the figures.

\end{acknowledgement}

\begin{suppinfo}
X-ray diffraction and X-ray reflectivity of NdNiO$_3$; X-ray reflectivity of the transducer layer; AFM of NdNiO$_3$; full derivation of the carrier and temperature model for FDPR signals, and data analysis. 
\end{suppinfo}

\bibliography{references}

\end{document}



\begin{tocentry}

    \includegraphics[width=7.5cm]{abstract.png}    
    \label{fig:toc}
\end{tocentry}

\newpage
\setcounter{equation}{0}
\setcounter{figure}{0}
\setcounter{table}{0}
\setcounter{page}{1}
\setcounter{section}{0}
\makeatletter
\renewcommand{\theequation}{S\arabic{equation}}
\renewcommand{\thefigure}{S\arabic{figure}}
\renewcommand{\thetable}{S\arabic{table}}

\renewcommand{\bibnumfmt}[1]{[S#1]}
\renewcommand{\citenumfont}[1]{S#1}
\renewcommand{\thesection}{S\arabic{section}}
\section{Methods}

\label{sec:Methods}
 
\subsection{Sample Preparation}
Thin-film NdNiO$_3$ was synthesized on LaAlO$_3$ (100) using ozone-assisted molecular-beam epitaxy (MBE)\cite{pan_synthesis_2022}. LaAlO$_3$ has a smaller lattice mismatch than other possible substrates, enabling the thickest film without relaxation.  During the deposition, the substrate temperature was kept at $\mathrm{555^\circ}$C and the chamber pressure was maintained at $\sim1.3\times 10^{-6}$ torr distilled ozone (Heeg Vacuum Engineering). We expect biaxial compressive strain of  0.4\%. The unit cell volume change across the transition is negligible. \textit{In-situ} reflection high-energy electron diffraction (RHEED), X-ray diffraction, and X-ray reflectivity measurements (PANalytical Empyrean X-Ray Diffractometer) demonstrate a highly crystalline, epitaxially strained sample with a smooth surface, with measured thickness 57.5 nm, approximately 150 unit cells (see Fig.~\ref{fig:rheed},~\ref{fig:xrd}). Fig.~\ref{fig:jampm160} details the synthesis and characterization of the 20.6 nm sample. We measured the sample resistance (Fig.~1) via a standard 4-probe AC resistance measurement in a physical property measurement system (PPMS).

To determine the thermal properties of the substrate, we coated a sample of LaAlO$_\mathrm{3}$ (Shinkosha) with a transducer layer. We deposited 56 nm Au with a 2.6 nm Cr adhesion layer and confirmed the thickness using x-ray reflectivity (PANalytical Empyrean X-Ray Diffractometer). We found that the Cr adhesion layer in the LaAlO$_\mathrm{3}$ sample created nonequilibrium electron-phonon coupling in the transducer layer and increased the phase lag for modulation frequencies above 10 MHz \cite{wilson_anisotropic_2014}; this effect is beyond the scope of our model. We still obtained accurate thermal parameters from lower modulation frequencies in the LaAlO$_\mathrm{3}$ sample, as the measurement of the substrate properties was sensitive to the phase at lower modulation frequencies (see Fig.~\ref{fig:elph},~\ref{fig:lao_sens}). 

We performed FDTR and FDPR measurements on the same sample of NdNiO$_3$ by coating half of the sample with a transducer layer. We deposited a transducer layer over a small region of the sample's surface, leaving much of the sample uncovered (see Fig.~2a). For the NdNiO$_3$ sample, we deposited 65.7 nm Au with a 1.8 nm Ti adhesion layer using electron-beam evaporation (Denton E-beam Evaporator). We confirmed the thickness with X-ray reflectivity as above. Based on the results from the substrate FDTR, we chose to use a Ti adhesion layer for the NdNiO$_3$ to eliminate transducer electron-phonon coupling. We could then model the phase at higher modulation frequencies where it is sensitive to the film's properties, as shown in Fig.~2c. The transducer we deposited on the NdNiO$_3$ film had a roughness of 14.16 \AA (Fig.~\ref{fig:afm}). These transducer layers were straightforward to model for our desired modulation frequencies, as gold is frequently used for FDTR, and its thin film thermal conductivity and heat capacity are known \cite{yang_modeling_2016}. The gold coating thermal conductivity was independently measured using the 4-point probe method (Fig.~\ref{fig:goldkappa}).

\subsection{FDTR and FDPR}
We use the FDTR and FDPR setup supplied by Fourier Scientific. To change the pump frequency, we modulate the voltage of a 458 nm 20 mW pump laser (Coherent OBIS LX fiber laser) and sweep the modulation frequency. The incident pump power is measured to be 17.3 mW. We measured the thermal conductivity at lower pump powers (Fig.~S9) and found similar values, and the lack of hysteresis persists. We also eliminate the possibility of a photo-induced MIT (Fig~S11). We record the amplitude and phase of $\Delta R(\omega)$ using a 532 nm 15 mW continuous-wave probe laser (Coherent OBIS LS laser) and a lock-in amplifier (Zurich Instruments HF2LI).

Both the pump and probe beams are nearly isotropic with Gaussian spatial profiles; the $1/e^2$ radius of the probe beam is 2.98 \textmu m. To minimize the noise in the probe signal, we rotate a half-wave plate in the probe beam path before it reaches the sample, which minimizes the difference in the intensity of the reference and reflected probe beams. Before we make a measurement, we measure the phase of the pump beam to subtract it from the measured phase. When measuring, we send the probe beams before and after reflection from the sample to a balanced photodetector. After sending the signal from the balanced detector to the lock-in amplifier\cite{yang_thermal_2013}, we obtain $\Delta R(\omega)$ for a range of frequencies $\omega$. We controlled the temperature using a flow cryostat (Instec HCP421V).

For substrate measurements using FDTR, we choose $C_p$ and $\kappa_z$ as the fitting variables in the fitting analysis. For thin‑film measurements, we choose $\kappa_\perp$ as the fitting variable for FDTR, and $D_a$ and $G_{FS}$ for FDPR. Specifically, we used the temperature-dependent $\kappa_\perp$ obtained from FDTR as an input parameter used in the FDPR modeling.
To extract the fitting variables of interest, we fit the phase of $\Delta R(\omega)$ using a nonlinear least-squares optimization to minimize the difference between the measured phase of $\Delta R(\omega)$ and the model prediction. 
Further details of the FDPR modeling formalism, along with additional parameters used in both FDTR and FDPR modeling, can be found below.

\section{Characterization of NdNiO$_3$ Film}

\subsection{Film Measurements}
\begin{figure}[h]
    \centering
    \includegraphics[width=\linewidth]{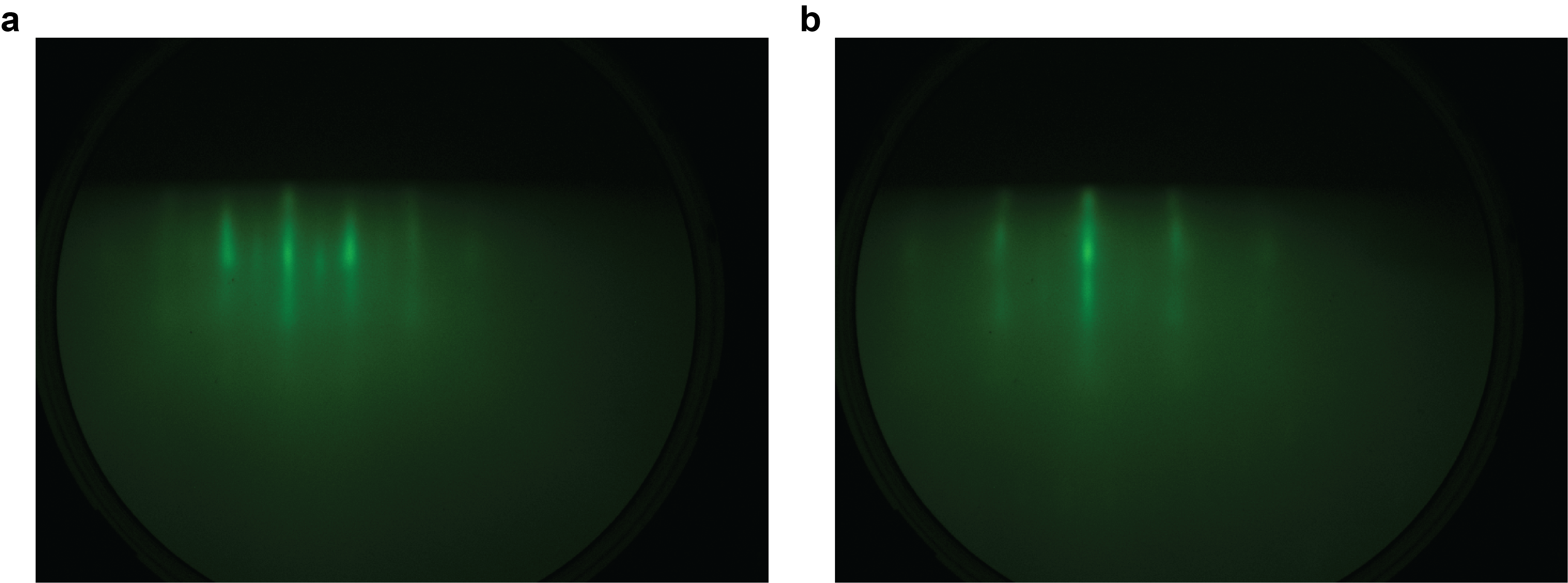}
    \caption{Reflection high-energy electron diffraction (RHEED) of NdNiO$_3$ on  LaAlO$_3$ during MBE growth. Patterns recorded in the \textbf{a}, (110) orientation and \textbf{b,} (100) orientation.}
    \label{fig:rheed}
\end{figure}

We measured the thickness and roughness of the film \textit{in situ} and after the film growth. We used reflection high-energy electron diffraction (RHEED) to characterize the film \textit{in situ}. Figure~\ref{fig:rheed} shows the diffraction pattern of the film during growth in the (100) and (110) orientations. The pattern reveals a smooth, epitaxial film. Figure~\ref{fig:xrd}a shows the film's X-ray diffraction pattern, taken after growth. The peaks are consistent with epitaxial NdNiO$_3$ in the (001) orientation. The best fit for the X-ray reflectivity curve (Fig.~\ref{fig:xrd}b) shows the film to be 57.5 nm thick with a roughness of 6.5 \AA. We examine the surface with atomic force microscopy (AFM), Fig.~\ref{fig:afm}a. Terracing from the substrate miscut and some defects in the surface are visible. We measured bulk electrical transport (Fig.~1) on a Quantum Design Physical Property Measurements System (PPMS) by applying 0.01 mA AC at 18.31 Hz via wire-bonded contacts .

\subsection{Coating Measurements}
\begin{figure}[h!]
    \centering
    \includegraphics[width=\linewidth]{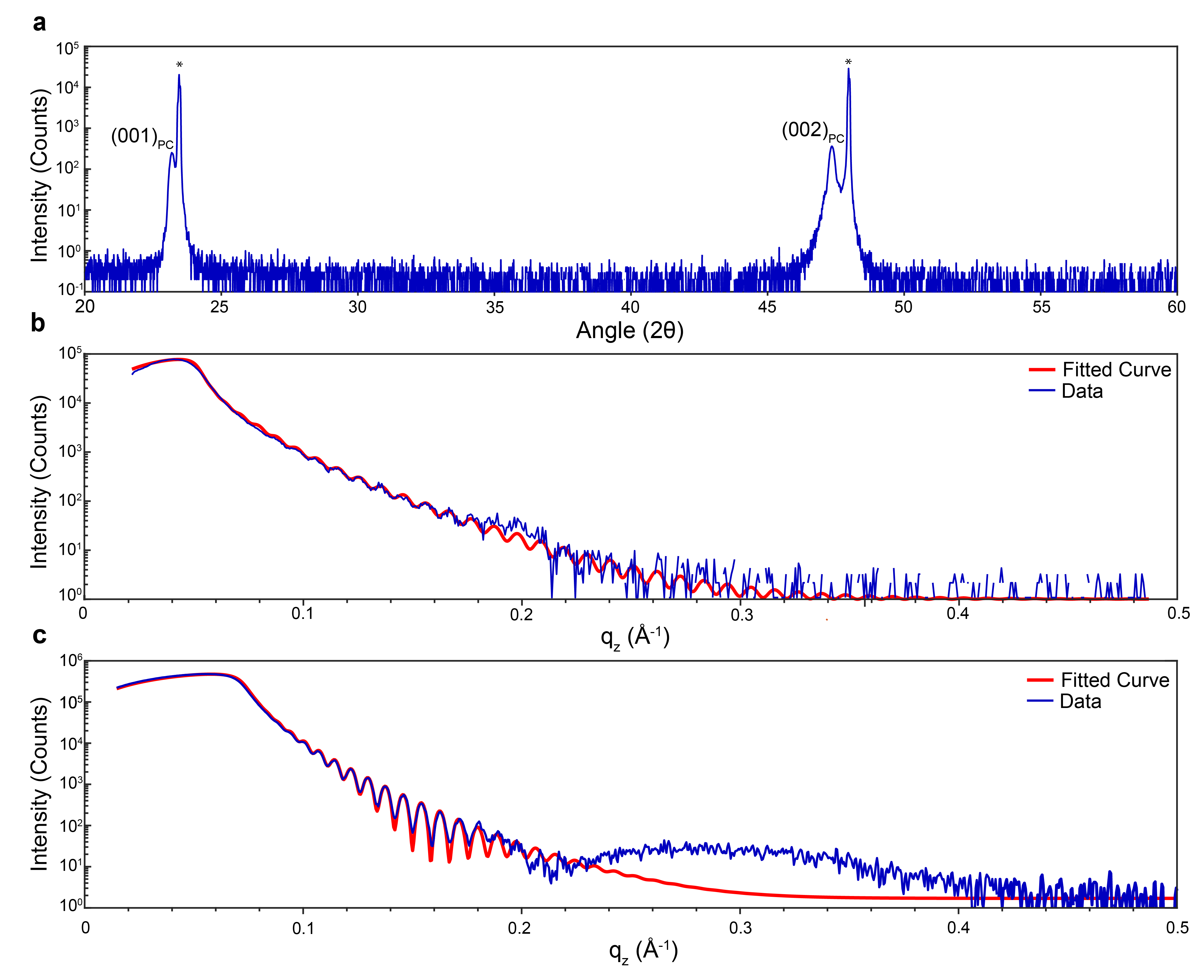}
    \caption{\textbf{a}, X-ray diffraction of 57.5 nm film showing epitaxial NdNiO$_3$ (001)$_\mathrm{PC}$ peaks. The asterisks denote LaAlO$_3$ (100) substrate peaks.  \textbf{b}, X-ray reflectivity of the 57.5 nm film after growth with the best fit overlaid. \textbf{c}, The X-ray reflectivity of the Au/Ti coating on the 57.5 nm film, with the best fit curve overlaid. The bump after the oscillations is due to the silicon sample holder.}
    \label{fig:xrd}
\end{figure}

\begin{figure}
    \centering
    \includegraphics[width=0.85\linewidth]{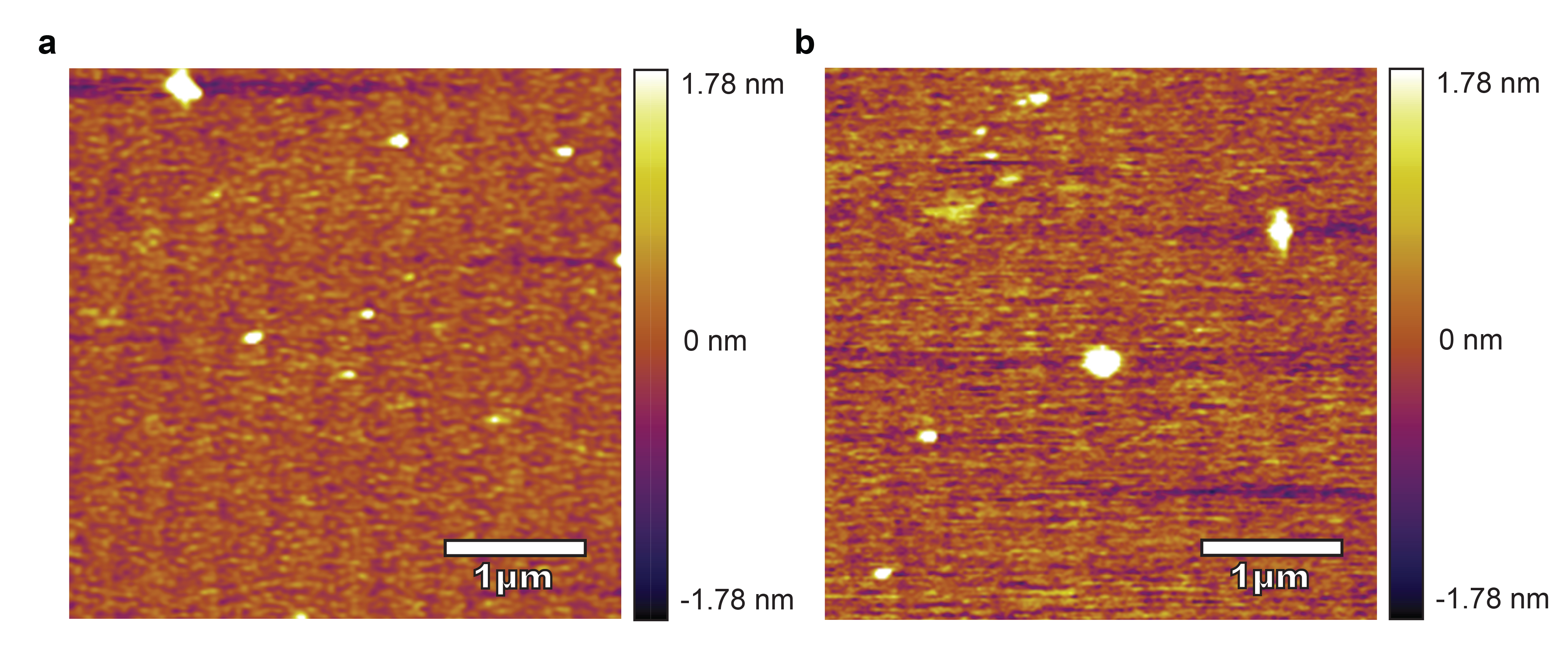}
    \caption{\textbf{a}, AFM characterization of 57.5 nm thick NdNiO$_3$ film showing defects (bright spots) along with terracing due to the substrate miscut (vertical striping). \textbf{b}, AFM characterization of the NdNiO$_3$ film after being coated. The film is slightly rougher overall, but the terracing is still visible, and there are still surface defects.}
    \label{fig:afm}
\end{figure}

The X-ray reflectivity (Fig.~\ref{fig:xrd}c) and AFM (Fig.~\ref{fig:xrd}b) of the gold-coated surface are also shown. The best fit for the X-ray reflectivity showed the coating to be 65.7 nm of gold and 1.8 nm of titanium, with a surface roughness of 14.16 \AA (Fig.~\ref{fig:xrd},\ref{fig:afm}). AFM confirms that the terracing is still visible, but the surface is slightly rougher overall, and there are still defects.

\clearpage

\clearpage
\newpage

\section{Mapping FDPR Phase Fluctuations}

\begin{figure}[h]
    \centering
    \includegraphics[width=0.9\linewidth]{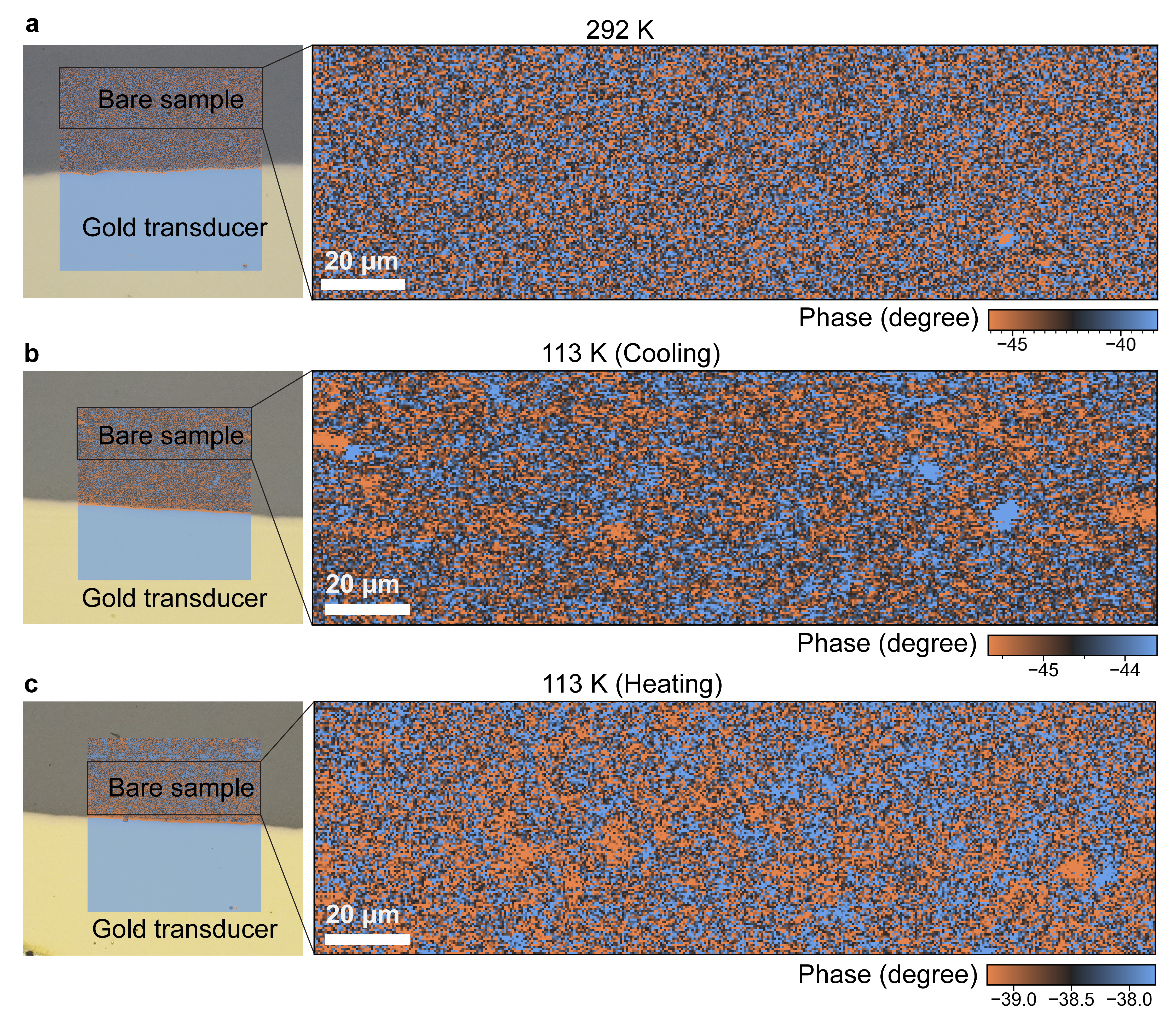}
    \caption{\textbf{a}, Room-temperature map of phase lag at a modulation frequency of $4.4\times10^6$ Hz, collected at 20$\times$ magnification, and the effective beam radius is approximately 1.5 \textmu m. 
    While we can record phase lag at room temperature with this smaller beam spot, its increased vibration sensitivity makes a 10x objective a better choice at lower temperatures under vacuum. \textbf{b}, Map of phase lag at $1.4\times 10^7$ Hz under 10x objective. \textbf{c}, Map of phase lag at $2.3\times10^6$ Hz under 10x objective. The large-scale fluctuations we observe at 113 K exist over a range of intermediate modulation frequencies on both cooling and heating, and they do not exist at any frequency in the room temperature map. }
    \label{fig:maps_all_rev2}
\end{figure}

Spatial FDPR mapping further reveals potential indications of phase coexistence near the transition, reinforcing a picture of spatially inhomogeneous, domain-driven phase evolution. \cite{bisht_phase_2017,preziosi_direct_2018,Gomes_structural_2024,lee_imaging_2019} We measured spatial FDPR maps at room temperature (Fig.~\ref{fig:maps_all_rev2}a), at $T_{\mathrm{switch}}$ on cooling (Fig.~\ref{fig:maps_all_rev2}b) and on heating (Fig.~\ref{fig:maps_all_rev2}c). At $T_{\mathrm{switch}}$, we observe longer length scale phase fluctuations that persist through a range of modulation frequencies and are present on both cooling and heating. 

We also recorded an FDPR phase map using a smaller beam size on the same sample at room temperature to measure the length scale of its fluctuations. We observe featureless, noise-induced fluctuations on a much smaller length scale at room temperature.  

\clearpage
\newpage

\section{Thermal Conductivity Data during Temperature Cycling}
\begin{figure}[h]
    \centering
    \includegraphics[width=0.7\linewidth]{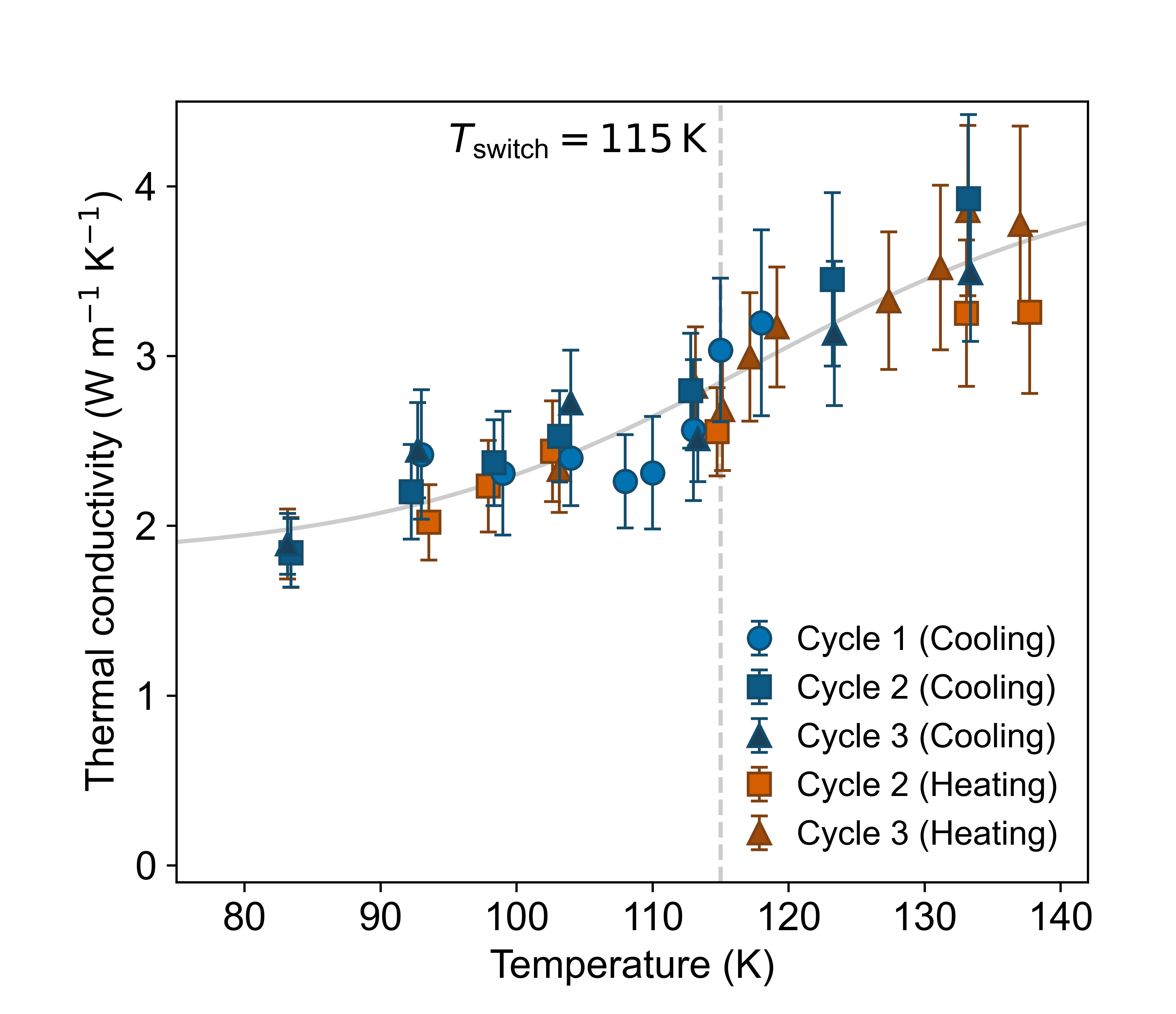}
    \caption{The thermal conductivity during heating and cooling cycles. The averaged values are presented in Fig.~3b.}
    \label{fig:kappa_unaveraged}
\end{figure}

The data in Fig.~3b were collected over 3 cycles for cooling and 2 cycles for heating and averaged. The data without averaging is shown in Fig.~\ref{fig:kappa_unaveraged}.

\clearpage
\section{Bootstrap Analysis for $ \boldsymbol{T_\mathrm{switch}}$}

\begin{figure}[h]
    \centering
    \includegraphics[width =0.59\columnwidth]{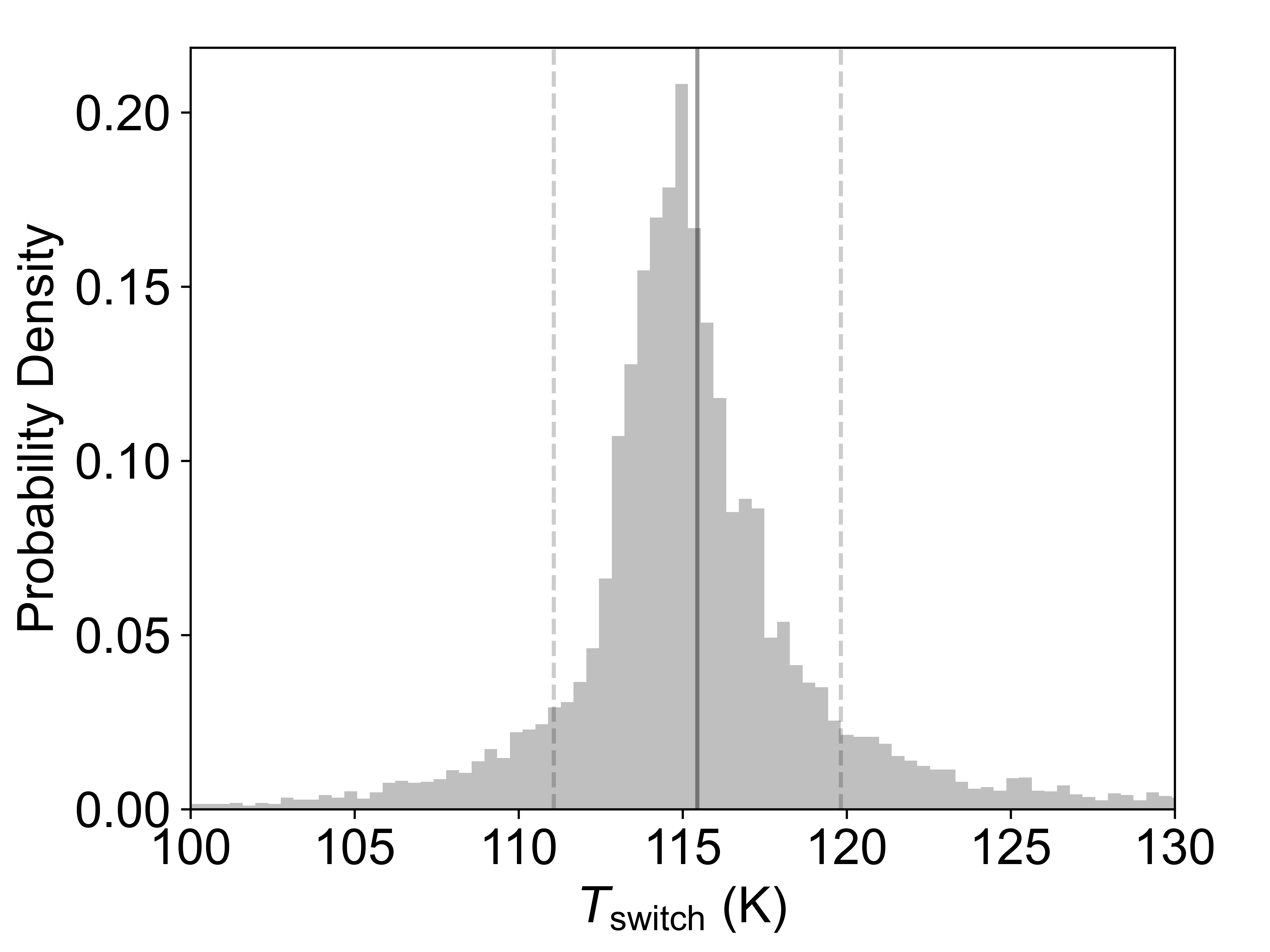}
    \caption{Histogram of  bootstrap‑derived $T_\mathrm{switch}$ values. Vertical lines mark the mean and $\pm1\sigma$ interval. The switching temperature is determined to be $T_\mathrm{switch}=115 \pm 4$ K.}
    \label{tswitch}
\end{figure}

We generate 10,000 synthetic datasets, sampling each data point from a normal distribution centered at its measured thermal conductivity value with a standard deviation given by its error bar. As shown in Fig.~\ref{tswitch}, the distribution of $T_\mathrm{switch}$ does not follow a perfect Gaussian function and exhibits slight asymmetry, potentially due to the slightly enlarged error bars at higher temperatures. We find the mean and standard deviation of $T_\mathrm{switch}$ is 115 K and 4 K.

\clearpage

\section{Uncertainty Analysis for Multivariate Fitting}

\begin{figure}
    \centering
\includegraphics[width=\linewidth]{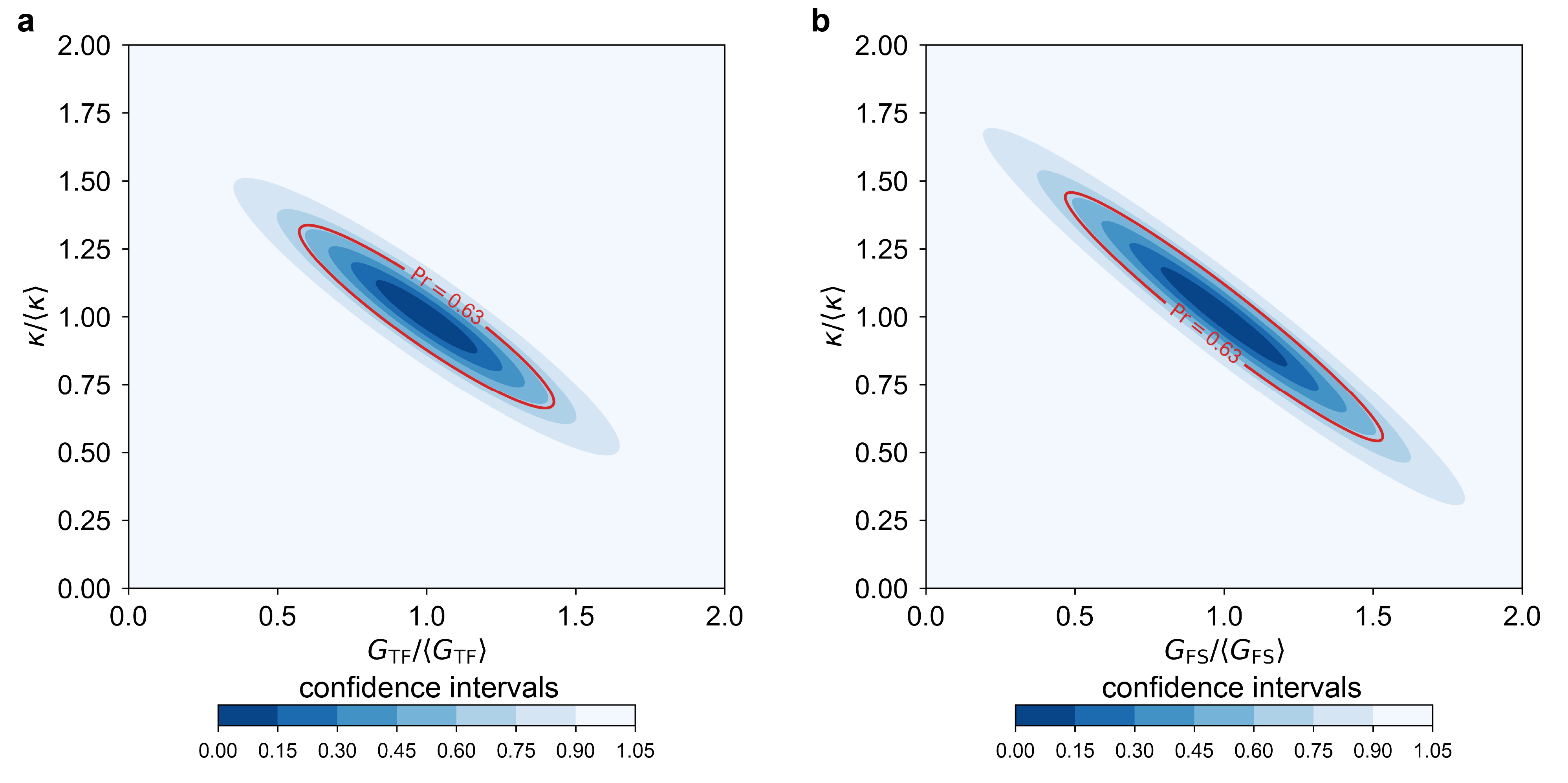}
    \caption{\textbf{a},\textbf{b}, The confidence intervals for fitting $\kappa$ and $G_\mathrm{TF}$ (\textbf{a}) and for fitting $\kappa$ and $G_\mathrm{FS}$ (\textbf{b}) at 110 K. The contours correspond to the cumulative distribution function of a chi-squared continuous random variable evaluated at the same Mahalanobis distance. The Mahalanobis distance is defined by $D=(\textbf{X}-\bm{\mu})^\mathrm{T}\bm{\Sigma}^{-1}(\textbf{X}-\bm{\mu})$, where $\textbf{X}$ is the random variable, $\bm{\mu}$ is the mean and $\bm{\Sigma}$ is the variance-covarience matrix. All non-fitting parameters are assumed to have a 3\% uncertainty. The 0.63 confidence interval, equivalent to the generalized 1-$\sigma$ confidence interval in the multivariate case, is marked in red. The bracket represents the mean value of the variable. We find that $\kappa$ and $G_\mathrm{TF}$ and $\kappa$ and $G_\mathrm{FS}$ are both anti-correlated, and that fitting the thermal conductivity of the thin film together with either thermal boundary conductance substantially increase the uncertainty in $\kappa$.}
    \label{fig:uncertainty}
\end{figure}

In processing the FDTR data with Au transducer, the heat transfer model involves the thermal boundary conductance between the metal transducer and the thin film, $G_\mathrm{TF}$, the thermal conductivity of the thin film, $\kappa$, and the thermal boundary conductance, $G_\mathrm{FS}$. The correlations between these variables prevent us from fitting them together. Here, we present the uncertainty analysis\cite{yang2016uncertainty} for fitting $\kappa$ and $G_\mathrm{TF}$ and for fitting $\kappa$ and $G_\mathrm{FS}$ by examining the confidence intervals in the multivariate fitting in Fig.~\ref{fig:uncertainty}. Because of the anti-correlations between $\kappa$ and $G_\mathrm{TF}$ and between $\kappa$ and $G_\mathrm{FS}$, we fix the values of both thermal boundary conductances ($G_\mathrm{TF}=G_\mathrm{FS}=45$ $\mathrm{MW\,m^{-2}\,K^{-1}}$) and only fit the thermal conductivity.

\clearpage

\begin{figure}[h]
    \centering
    \includegraphics[width =0.9\columnwidth]{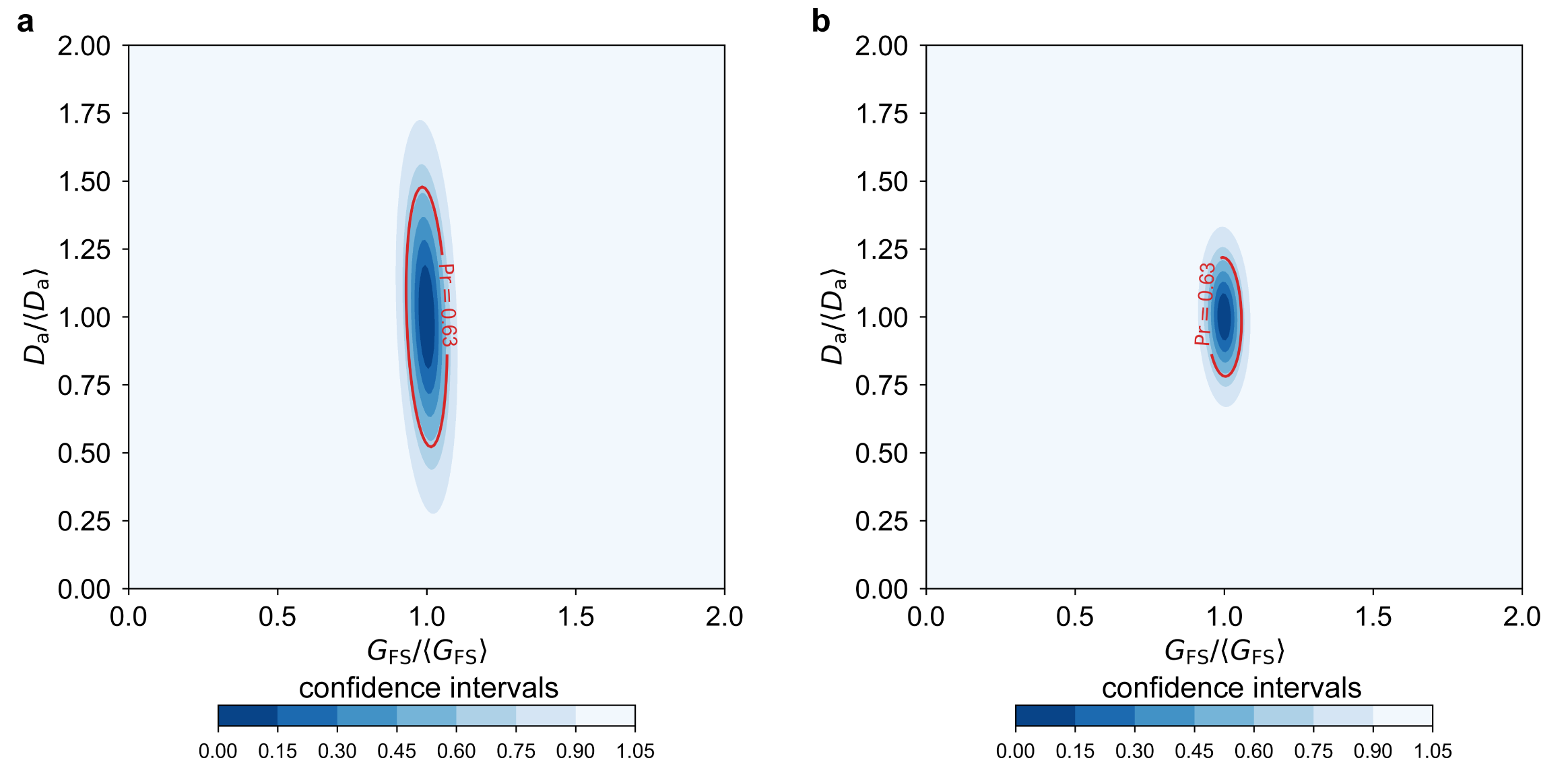}
    \caption{\textbf{a},\textbf{b}, The confidence intervals for extracting $G_\mathrm{FS}$ and $D_\mathrm{a}$ from FDPR data at 108 K (\textbf{a}) and 133 K (\textbf{b}). Both contour plots reveal minimal correlation between $G_\mathrm{FS}$ and $D_\mathrm{a}$.}
    \label{fdpr_cov}
\end{figure}

In processing the FDPR phase data, we choose $D_\mathrm{a}$ and $G_\mathrm{FS}$ as the fitting variables. This is because the sensitivity curves for $D_\mathrm{a}$ and $G_\mathrm{FS}$ possess distinctive frequency dependence (i.e. they are not simple rescalings of one another) in Fig.~2. We also calculate the variance-covarience matrix and present the corresponding confidence intervals for fitting $D_\mathrm{a}$ and $G_\mathrm{FS}$ in Fig.~\ref{fdpr_cov}. The  confidence intervals reveal essentially zero correlation, further  supporting our fitting strategy for the FDPR data analysis.

\clearpage
\section{Effect of Pump Power on the Microscopic Domain Effect}

\begin{figure}
    \centering
    \includegraphics[width=0.5\linewidth]{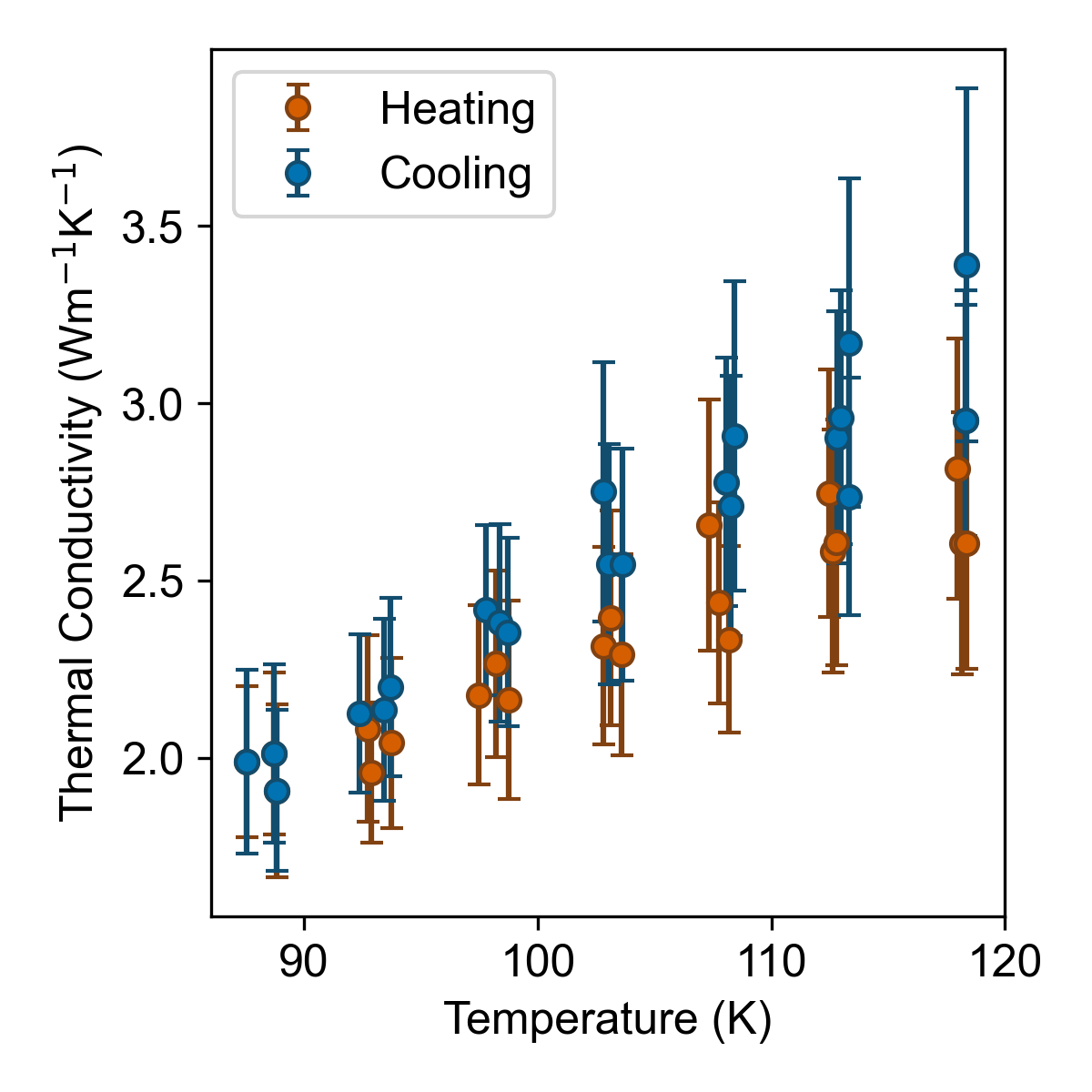}
    \caption{To eliminate the possibility of the high pump power reducing the observed hysteresis in Fig.~3b, we measured the thermal conductivity using an incident pump power of 5 mW and a probe power of 10 mW on heating and cooling. The temperature rise for this pump power is expected to be less than 6 K. We still observe little hysteresis between heating and cooling, and the thermal conductivity values are consistent with those from the higher power measurement in Fig.~3b.}
    \label{fig:kappa_lowpower}
\end{figure}

\section{Characterization of 20.6 nm sample}
\begin{figure}
    \centering
    \includegraphics[width=\linewidth]{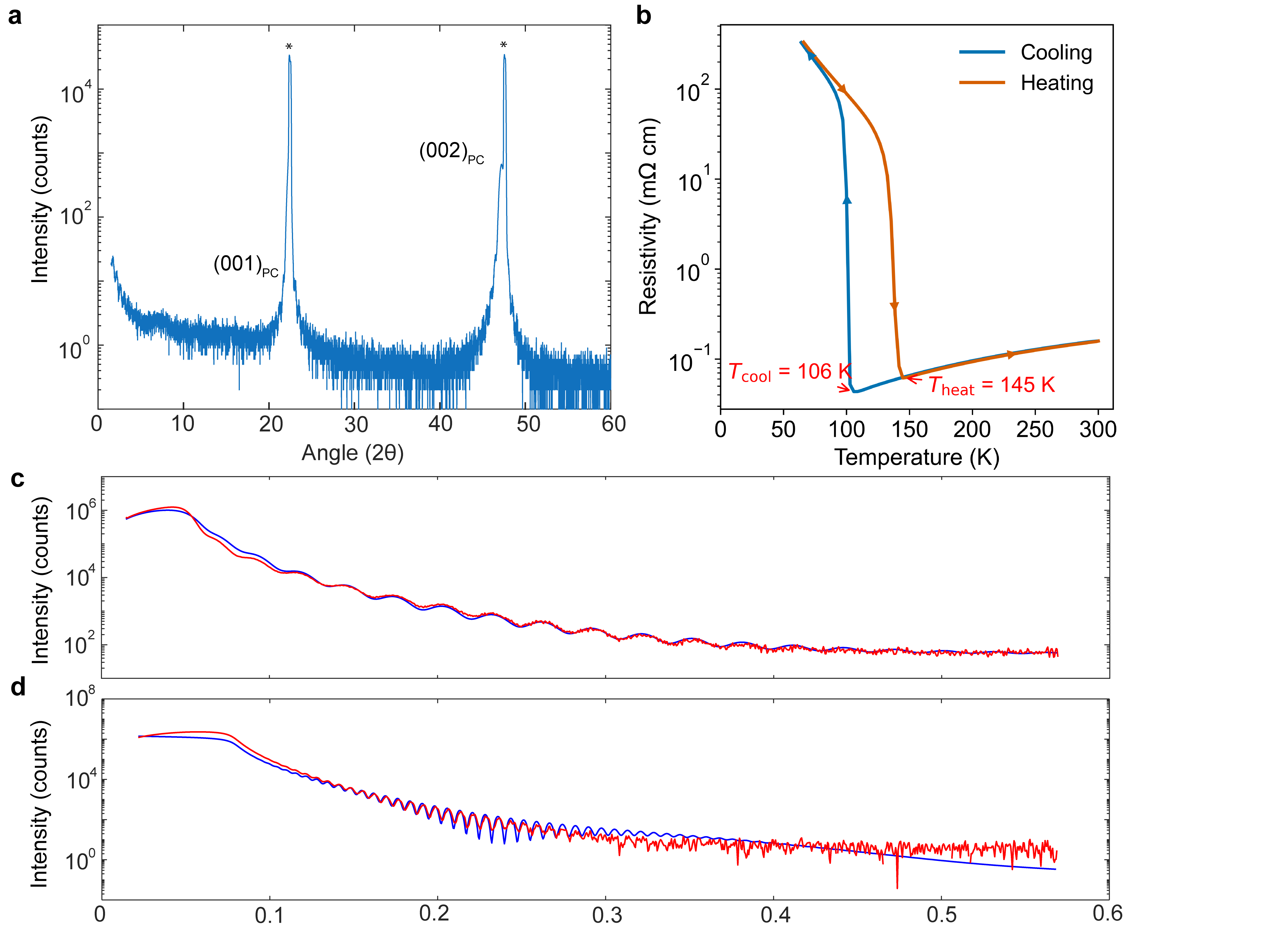}
    \caption{Characterization of 20.6-nm NdNiO$_3$ thin film 
    on LaAlO$_3$ substrate. \textbf{a}, X-ray diffraction pattern of the sample showing the pseudocubic (001) and (002) peaks. LaAlO$_3$ (001)$_\text{PC}$ and (002)$_\text{PC}$ peaks are marked with an asterisk (*). \textbf{b}, Temperature-dependent electrical resistivity showing T$_\mathrm{MIT}$ = 106 K on cooling and 145 K on heating, yielding a hysteresis width of 39 K. \textbf{c}, X-ray reflectivity data on the sample (red), with the best fit for a 20.6 nm thickness overlaid (blue). \textbf{d}, X-ray reflectivity data (blue) for the transducer layer with the best fit (red) overlaid, corresponding to 77.8 nm of Au and 1.8 nm of Ti.}
    \label{fig:jampm160}
\end{figure}

We measure a 20.6 nm NdNiO$_3$ thin film on LaAlO$_3$ substrate grown via molecular-beam epitaxy. Four-probe resistivity measurements (Fig.~\ref{fig:jampm160}b) show $T_\mathrm{MIT}=$ 106 K on cooling and $T_\mathrm{MIT}=$ 145 K on warming; both temperatures are slightly higher than in the original sample, and the hysteresis of 39 K is comparable to the original sample. We deposit a transducer layer of 77.8 nm Au and a 1.8 nm Ti adhesion layer via electron-beam evaporation (Denton e-beam evaporator). This procedure is consistent with that used for our original sample, expect that we used a Cr adhesion layer only on the LaAlO$_3$ substrate sample. 
We confirm the thickness of the film and the transducer layer, as well as the film's composition, using X-ray reflectivity (Fig.~\ref{fig:jampm160}c,d) and X-ray diffraction (Fig.~\ref{fig:jampm160}a) using a PANalytical Empyrean X-Ray Diffractometer. We measure the thermal conductivity across the MIT during heating and cooling to identify any hysteresis. 
As in the low power measurements in Fig.~\ref{fig:kappa_lowpower}, we measure this sample using an incident pump power of 5 mW. 

\clearpage
\newpage
\clearpage

\clearpage

\section{Refractive Index of NdNiO$_3$}
According to Ref.~\cite{katsufuji_optical_1995}, the real part of the dielectric constant of bulk NdNiO$_3$ is approaching a constant value of $\varepsilon_r$ = 1.7 as the photon energy $\omega$ approaches 1 eV. This value remains nearly unchanged across temperatures both below and above the metal–insulator transition temperature. A similar asymptotic behavior has been observed in NdNiO$_3$/NdGaO$_\mathrm{3}$ (110)~\cite{ruppen_optical_2015}. Therefore, we approximate the real part of the dielectric constant of the NdNiO$_3$ thin film with $\varepsilon_r$ = 1.7 at photon energies of $\omega$ = 2.71 eV (pump wavelength $\lambda$ = 458 nm) and $\omega$ = 2.33 eV (probe wavelength $\lambda$ = 532 nm). As for the imaginary part of the dielectric constant, the optical conductivity measurements on a NdNiO$_3$ film on a LaAlO$_\mathrm{3}$ (001) substrate in Ref.~\cite{stewart_mott_2011} indicates that $\varepsilon_i$ = 2.86 at $\omega$ = 2.71 eV and $\varepsilon_i$ = 3.15 at $\omega$ = 2.33 eV at 298 K. Notably, the imaginary dielectric constant between 2 eV and 3 eV varies minimally across temperatures below and above the metal–insulator transition. Hence, we obtain that $\varepsilon_i$ = 2.86 at the pump photon energy and $\varepsilon_i$ = 3.15 at the probe photon energy. 
Accordingly, the real and imaginary parts of the refractive index are $n$ = 1.59, $\kappa$ = 0.90 at the pump wavelength, and $n$ = 1.59, $\kappa$ = 0.97 at the probe wavelength. 
Consequently, the optical penetration depths for the pump and probe beams are 40 nm and 43 nm, respectively. 

\clearpage

\section{Frequency-Domain Photo-Reflectance for the Thin-Film-on-Substrate Geometry}

The frequency-domain photo-reflectance signal is contributed by the carrier density part, the electronic temperature part, and the lattice temperature part,
\begin{equation}
    \frac{\Delta R}{R}=\mathrm{Re}\{A\overline{\rho}+B\overline{T}_{\mathrm{el}}+C\overline{T}_{\mathrm{ph}}\}
    \label{eq:fdtr1}
\end{equation}
where $\overline{\phantom{x}}$ denotes the weighted average by the probe beam profile, and $A$, $B$, and $C$ are material-dependent real constants.
As pointed out by Ref.~\cite{song_probing_2024}, in the modulation frequency range we choose ($<$ 10 MHz), the electron-phonon coupling is strong enough such that $T_{\mathrm{el}}\approx T_{\mathrm{ph}}$. As a result, we only need to consider the contributions from the carrier and the lattice temperature, 
\begin{equation} 
\label{eq:ccr_ctr}
\frac{\Delta R}{R} \approx \mathrm{Re} \{ \mathrm{CCR} \;\overline{\rho} + \mathrm{CTR}\; \overline{T} \} \end{equation}
where $\mathrm{CCR}$ is the coefficient of carrier-induced reflectance, and $\mathrm{CTR} = B + C$ is the effective coefficient of thermoreflectance. In the following, we outline how to model the signal due to carriers and temperature.

We start with the dynamics of photo-excited carriers in a thin film described by,
\begin{equation}
    \frac{d\rho}{dt} = D_{\mathrm{a},z}  \frac{\partial^2\rho}{\partial z^2} + D_{\mathrm{a},\parallel} \left(\frac{\partial^2}{\partial x^2}+\frac{\partial^2}{\partial y^2}\right)\rho- \frac{\rho}{\tau} + P
    \label{carrier}
\end{equation}
where $\rho$ is the excited electron density, $P$ is the number density of pump photons absorbed by the thin film per unit time, $D_{\mathrm{a},\parallel}$ and $D_{\mathrm{a},z}$ are the in-plane and out-of-plane ambipolar diffusivity, respectively, and $\tau$ is the carrier recombination time. Taking the Fourier transform of Eq.~\ref{carrier} in the x- and y-directions, we obtain,
\begin{equation}
    \frac{\partial^2}{\partial z^2}\tilde{\rho} -\Lambda \tilde{\rho}+\frac{\tilde{P}}{D_{\mathrm{a},z}} = 0
    \label{carrier-ft}
\end{equation}
where $\tilde{P}=\frac{P_0\alpha}{h\nu_\mathrm{pump}}e^{-\frac{\sigma_x^2q_x^2}{8}}e^{-\frac{\sigma_y^2q_y^2}{8}}$ and $\Lambda = \frac{D_{\mathrm{a},\parallel}}{D_{\mathrm{a},z}}q_\parallel^2 + \frac{1}{D_{\mathrm{a},z}\tau}+\frac{i\omega}{D_{\mathrm{a},z}}$.
Here, $P_0$ is the absorbed pump laser power, $h\nu_\mathrm{pump}$ is the photon energy, $\alpha$ is the inverse penetration depth of the pump laser beam, and $\sigma_x$ and $\sigma_y$ are the pump laser beam's radii along the x- and y-directions, respectively. 
The boundary condition at the two surfaces of the film is given by,
\begin{equation}
    -D_{\mathrm{a},z}\frac{\partial \tilde{\rho}}{\partial z} +\tilde{\rho} S_j = 0
\end{equation}
where $S_j$ $(j=1,2)$ denotes the surface recombination velocities at the top and bottom surface, respectively. Note that the bottom surface corresponds to the interface between the thin film and the insulating substrate.

The solution to Eq.~\ref{carrier-ft} takes the following form,
\begin{equation}
    \tilde{\rho} = De^{-\sqrt{\Lambda}z}+Ee^{\sqrt{\Lambda}z}+Fe^{-\alpha z}
\end{equation}
where $F = \frac{P_0\alpha}{h\nu_\mathrm{pump}{D_{\mathrm{a},z}(\Lambda-\alpha^2)}}e^{-\left(\sigma_xq_x^2+\sigma_yq_y^2\right)/8}$ and,
\begin{equation}
    D = \frac{-(D_{\mathrm{a},z}\alpha+S_1)(-D_{\mathrm{a},z}\sqrt{\Lambda}+S_2)+(-D_{\mathrm{a},z}\sqrt{\Lambda}+S_1)(D_{\mathrm{a},z}\alpha+S_2)e^{-(\sqrt{\Lambda}+\alpha)L}}{(D_{\mathrm{a},z}\sqrt{\Lambda}+S_1)(-D_{\mathrm{a},z}\sqrt{\Lambda}+S_2)-(-D_{\mathrm{a},z}\sqrt{\Lambda}+S_1)(D_{\mathrm{a},z}\sqrt{\Lambda}+S_2)e^{-2\sqrt{\Lambda}L}}F
\end{equation}
\begin{equation}
    E = \frac{-(D_{\mathrm{a},z}\alpha+S_1)(D_{\mathrm{a},z}\sqrt{\Lambda}+S_2)+(D_{\mathrm{a},z}\sqrt{\Lambda}+S_1)(D_{\mathrm{a},z}\alpha+S_2)e^{(\sqrt{\Lambda}-\alpha)L}}{(-D_\mathrm{a}\sqrt{\Lambda}+S_1)(D_{\mathrm{a},z}\sqrt{\Lambda}+S_2)-(D_{\mathrm{a},z}\sqrt{\Lambda}+S_1)(-D_{\mathrm{a},z}\sqrt{\Lambda}+S_2)e^{2\sqrt{\Lambda}L}}F
\end{equation}
with $L$ denoting the thickness of the thin film.

\begin{figure}[h!]
    \centering
    \includegraphics[width=\linewidth]{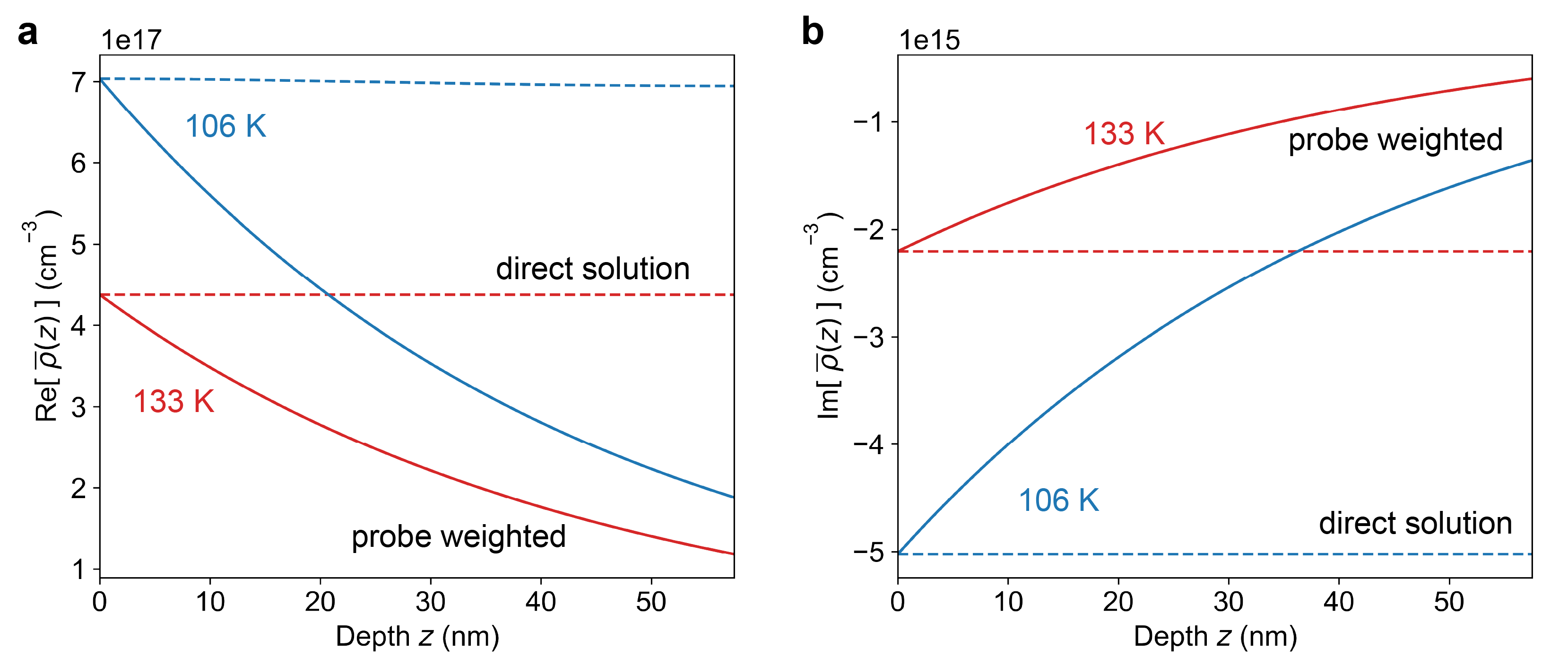}
    \caption{\textbf{a},\textbf{b}, Real and imaginary parts of the photo-excited carrier density $\overline{\rho}(z)$ within the thin film at $f=10.5$ MHz when the absorbed pump power $P_0 = 6$ mW. Solid lines represent the direct spatial profile $\overline{\rho}(z)$, while the dashed lines correspond to the carrier density weighted by the exponentially decaying probe intensity in the depth direction ($\propto e^{-\gamma z}$). The carrier part of the signal collected by the probe is given by $\mathrm{CCR}\times\overline{\rho} = \mathrm{CCR}\,\gamma\int\overline{\rho}(z)e^{-\gamma z}dz$. We rule out the possibility of a direct photo-induced metal-insulator transition due to the insufficient density of photo-excited carriers. Instead, the metal-insulator is temperature-driven. }
    \label{fig:SI_carrierdensity}
\end{figure}

We then take the inverse Fourier transform to obtain $\rho(\mathbf{r})$ and the signal measured by the probe beam is given by,
\begin{equation}I_\mathrm{e}(\omega)=\mathrm{CCR}\frac{2\gamma}{\pi\sigma_x^{'}\sigma_y^{'}}\int\int_0^\infty\rho(\mathbf{r})e^{-2x^2/\sigma_x^{'2}}e^{-2y^2/\sigma_y^{'2}}e^{-\gamma z}dzd^2\mathbf{r}_\parallel
\end{equation}
where $\gamma$ is the inverse penetration depth of the probe beam, and $\sigma_x^{'}$ and $\sigma_y^{'}$ are the probe beam radii along x- and y-directions, respectively. Note that we extend the upper limit of the integration in the 
z-direction to infinity, as the light entering the substrate can be ignored given the fact that both penetration depths of the pump and probe beams are smaller than the film thickness. In oxide thin films, such as VO$_\mathrm{2},$\cite{qazilbash_mott_2007} the electron mean free path is on the order of the lattice constant, and thus much shorter than the film thickness. Hence, we can assume isotropic diffusivity, such that $D_{\mathrm{a},\parallel}=D_{\mathrm{a},z}=D_\mathrm{a}$, and $\Lambda = q^2_\parallel + \frac{1}{D_\mathrm{a}\tau}+\frac{i\omega}{D_\mathrm{a}}$.
Further assuming small surface recombination velocities (weak band bending), we have the following expression of the signal,
\begin{equation}
    I_\mathrm{e}(\omega)= \frac{\mathrm{CCR}}{(2\pi)^2}\int \frac{P_0\alpha\gamma}{h\nu D_\mathrm{a}(\Lambda-\alpha^2)}\left(\frac{1}{\alpha+\gamma}+\frac{D}{\sqrt{\Lambda}+\gamma}+\frac{E}{-\sqrt{\Lambda}+\gamma}\right)e^{-\frac{R_x^2q^2_x}{4}}e^{-\frac{R_y^2q^2_y}{4}}d^2\mathbf{q}_\parallel
    \label{full_e}
\end{equation}
where
\begin{equation}
    D=\frac{\alpha}{\sqrt{\Lambda}}\frac{1-e^{-(\sqrt{\Lambda}+\alpha)L}}{-1+e^{-2\sqrt{\Lambda}L}}
\end{equation}
\begin{equation}
E = \frac{\alpha}{\sqrt{\Lambda}}\frac{-1+e^{(\sqrt{\Lambda}-\alpha)L}}{-1+e^{2\sqrt{\Lambda}L}}
\end{equation}
and $R_x = \sqrt{\sigma_x^2+\sigma_x^{'2}}$ and $R_y = \sqrt{\sigma_y^2+\sigma_y^{'2}}$ are the effective beam radii along x- and y-directions, respectively. Eq.~\ref{full_e} is the formal expression of the carrier part of the signal. In Fig.~\ref{fig:SI_carrierdensity}, we present the representative spatial profiles of the carrier density. The probe-intensity-weighted integral of this distribution corresponds to the carrier part of the measured signal.

Next, we consider the temperature part of the signal $I_\mathrm{th}(\omega)$. As derived in Ref.~\cite{yang_modeling_2016}, the phonon temperature profile within the thin film is expressed by,
\begin{equation}
    \tilde{T} = \left(Ce^{\sqrt{\lambda}z}+De^{-\sqrt{\lambda}z}+Fe^{-\alpha z}\right)e^{-\frac{\sigma_x^2q_x^2}{8}}e^{-\frac{\sigma_y^2q_y^2}{8}}
\end{equation}
where $\lambda=i\frac{C_p\omega}{\kappa_\perp}+\frac{\kappa_\parallel}{\kappa_\perp}q^2_\parallel$, $F=\frac{1}{\kappa_\perp}\frac{P_0\alpha}{\lambda-\alpha^2}$, and $C_p$ is the volumetric heat capacity. Here, $C$ and $D$ are given by,
\begin{equation}
    C =\frac{1}{2}\left[\frac{\alpha}{\sqrt{\lambda}}-\left(1+\frac{M_4}{M_3}\right)\right]F
\end{equation}
\begin{equation}
    D =\frac{1}{2}\left[-\frac{\alpha}{\sqrt{\lambda}}-\left(1+\frac{M_4}{M_3}\right)\right]F
\end{equation}
Specifically, $M_3$ and $M_4$ are elements of the transfer matrix relating the temperature and heat flux at the bottom of the substrate to the temperature and the laser-heating-related term at the surface,
\begin{equation}
    \begin{pmatrix}
    \tilde{T}_n(L_n)\\
    \tilde{q}_n(L_n)
    \end{pmatrix}
    =
    \begin{pmatrix}
        M_1&M_2\\
        M_3&M_4
    \end{pmatrix}
    \begin{pmatrix}
        \tilde{T}_1(0)\\
        Fe^{-\frac{R_x^2q^2_x}{4}}e^{-\frac{R_y^2q^2_y}{4}}
    \end{pmatrix}
\end{equation}
The construction of such a transfer matrix requires the heat capacity $C_p$, thickness $L$, and thermal conductivity components $\kappa_\parallel$ and $\kappa_\perp$ of each layer, as well as the thermal boundary conductance $G$ between adjacent layers.

Consequently, the temperature part of the signal is written as,
\begin{equation}
\begin{aligned}
    &I_\mathrm{th}(\omega)  = \frac{\mathrm{CTR}\,\gamma}{(2\pi)^2}\int \left(\frac{C}{\gamma-\sqrt{\lambda}}+\frac{D}{\sqrt{\lambda}+\gamma}+\frac{F}{\alpha+\gamma}\right)e^{-\frac{R_x^2q^2_x}{4}}e^{-\frac{R_y^2q^2_y}{4}}d^2\mathbf{q}_\parallel\\
    &=\frac{\mathrm{CTR}}{(2\pi)^2}\int \left\{ \frac{1}{\lambda-\gamma^2}\left[\left(1+\frac{M_4}{M_3}\right)\gamma-\alpha\right]+\frac{1}{\alpha+\gamma}\right\}\frac{P_0\alpha\gamma}{\kappa_\perp(\lambda-\alpha^2)}e^{-\frac{R_x^2q^2_x}{4}}e^{-\frac{R_y^2q^2_y}{4}}d^2\mathbf{q}_\parallel
    \end{aligned}
    \label{ph}
\end{equation}
Finally, by summing Eq.~\ref{full_e} and Eq.~\ref{ph}, we obtain the total signal measured at different modulation frequencies in the experiment,
\begin{equation}
    I(\omega) = I_\mathrm{e}(\omega) +  I_\mathrm{th}(\omega)
    \label{total}
\end{equation}
We refer to the combined expression in Eq.~\ref{total} as the carrier and temperature model.
We emphasize that, to the best of our knowledge, such a parallel treatment of charge and thermal transport in a thin film under laser irradiation has not been previously reported.
\clearpage
\clearpage
\begin{figure}[h!]
    \centering
    \includegraphics[width=0.7\linewidth]{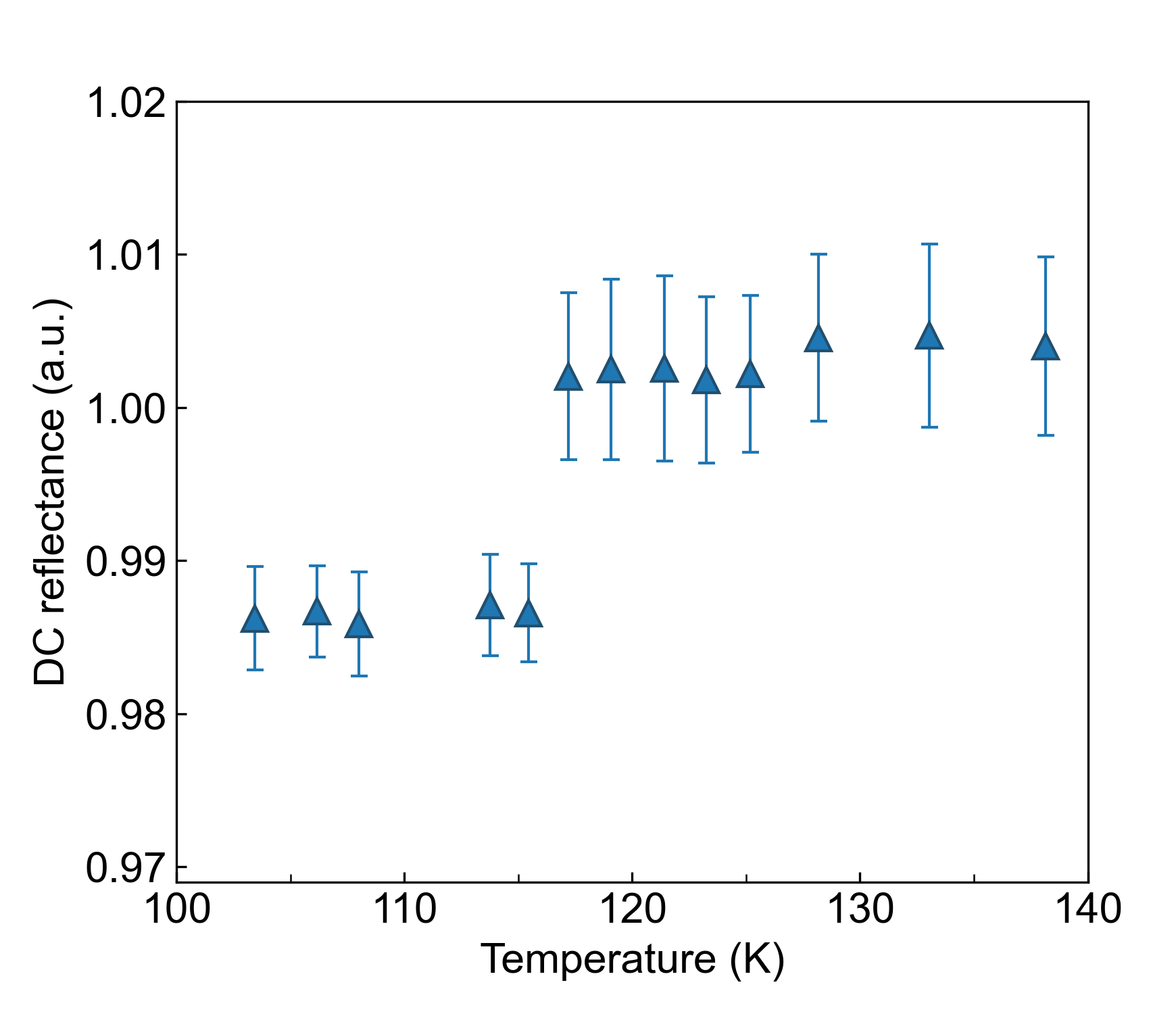}
    \caption{The DC photo-reflectance $R$ as a function of temperature in FDPR measurements during heating. The error bars represent the standard deviations of the data across all modulation frequencies at the same temperature and indicate only a weak frequency dependence. A sudden jump in $R$ is observed around 115 K, consistent with the transport measurements. }
    \label{fig:SI_reflectance}
\end{figure}
\clearpage
\section{Fitting Analysis in FDPR}

The total signal in Eq.~\ref{total} can be reorganized as,
\begin{equation}
    I(\omega) = \mathrm{CTR}\left(\overline{T}+\frac{\mathrm{CCR}}{\mathrm{CTR}}\times\overline{\rho}\right)
\end{equation}
In fitting the phase data of the photo-reflectance to our model, the prefactor $\mathrm{CTR}$ does not influence the best, and the combined term $\frac{\mathrm{CCR}}{\mathrm{CTR}}$ effectively behaves as an independent variable. Therefore, we define,
\begin{equation}
    A_\rho = \frac{\mathrm{CCR}}{\mathrm{CTR}}
\end{equation}
By considering Eq.~\ref{eq:ccr_ctr}, we identify the physical meaning of $A_\rho$ as the inverse of the temperature derivative of the carrier density,
\begin{equation}
    A_\rho=\frac{\Delta T}{\Delta \rho}\frac{}{}\approx\left(\frac{\partial\rho}{\partial T}\right)^{-1}
\end{equation}

In practice, fitting both the amplitude and phase of the photoreflectance signals allows us to extract the specific values of CCR and CTR. We have demonstrated this capability in prior work\cite{song_probing_2024} for bulk semiconductors. However, because of the strong anti-correlation between coefficient $A_\rho$ and ambipolar diffusivity $D_\mathrm{a}$ in the nanoscale thin-film-on-substrate geometry, we cannot accurately determine $A_\rho$. Even so, this does not affect the relative change observed for $D_\mathrm{a}$, irrelevant to the specific choice of $A_\rho$.

As a result, we need 3 unknowns to evaluate the carrier part of the signal: the ambipolar diffusion coefficient $D_\mathrm{a}$, the inverse of the temperature derivative of the carrier density $A_\rho$, and the carrier recombination lifetime $\tau$. The sensitivity to any variable $x_i$ at the $j$th modulation frequency $\omega_j$ can be characterized by the phase difference,
\begin{equation}
    \Delta \phi (\omega_j,x_i) =\mathrm{arctan}\left(\frac{\mathrm{Im}\{I(\omega_j,x_i+\Delta x_i)\}}{\mathrm{Re}\{I(\omega_j,x_i + \Delta x_i)\}}\right) -\mathrm{arctan}\left(\frac{\mathrm{Im}\{I(\omega_j,x_i)\}}{\mathrm{Re}\{I(\omega_j,x_i)\}}\right)
\end{equation}
Here, we take the change in the variable, $\Delta x_i$, to be 1\% of $x_i$, that is, $\Delta x_i = 0.01 x_i$.

As illustrated in Fig.~2, the sensitivities to $D_\mathrm{a}$ and $\tau$ are similar in absolute magnitude but with opposite signs. This suggests that $D_\mathrm{a}$ and $\tau$ are anti-correlated in fitting. The sensitivities with respect to $D_\mathrm{a}$ and $A_\rho$ exhibit a similar frequency dependence, differing only by a scaling factor on the order of unity. This indicates that $D_\mathrm{a}$ and $A_\rho$ are positively correlated in fitting. As a consequence, we can fit only one out of the three variables with confidence. We argue that $A_\rho$ and $\tau$ are more intimately connected to the band structure, which, within the energy range relevant to pump-excited electrons changes only minimally across the metal-insulator transition, whereas $D_\mathrm{a}$ is more strongly influenced by electron transport governed by electron–electron and electron–phonon interactions that are susceptible to metal-insulator transition. Hence, we treat $A_\rho$ and $\tau$ as constants, and choose $D_\mathrm{a}$ as a fitting variable. Note that the thermal boundary conductance $G_{FS}$ between the thin film and substrate and $D_\mathrm{a}$ are not correlated (see Fig.~\ref{fdpr_cov}), and thus we include $G_{FS}$ as an additional fitting parameter.
\clearpage
\section{Sensitivity Analysis of LaAlO$\mathbf{_3}$ Substrate}
We find that our FDTR measurements are highly sensitive to the substrate parameters shown in Fig.~3a, measured on a substrate sample with no film. We show below the change in phase for a 1\% increase in each of the shown parameters in the same way as calculated in Fig.~2d,e. 
\begin{figure}[h!]
    \centering
    \includegraphics[width=0.7\linewidth]{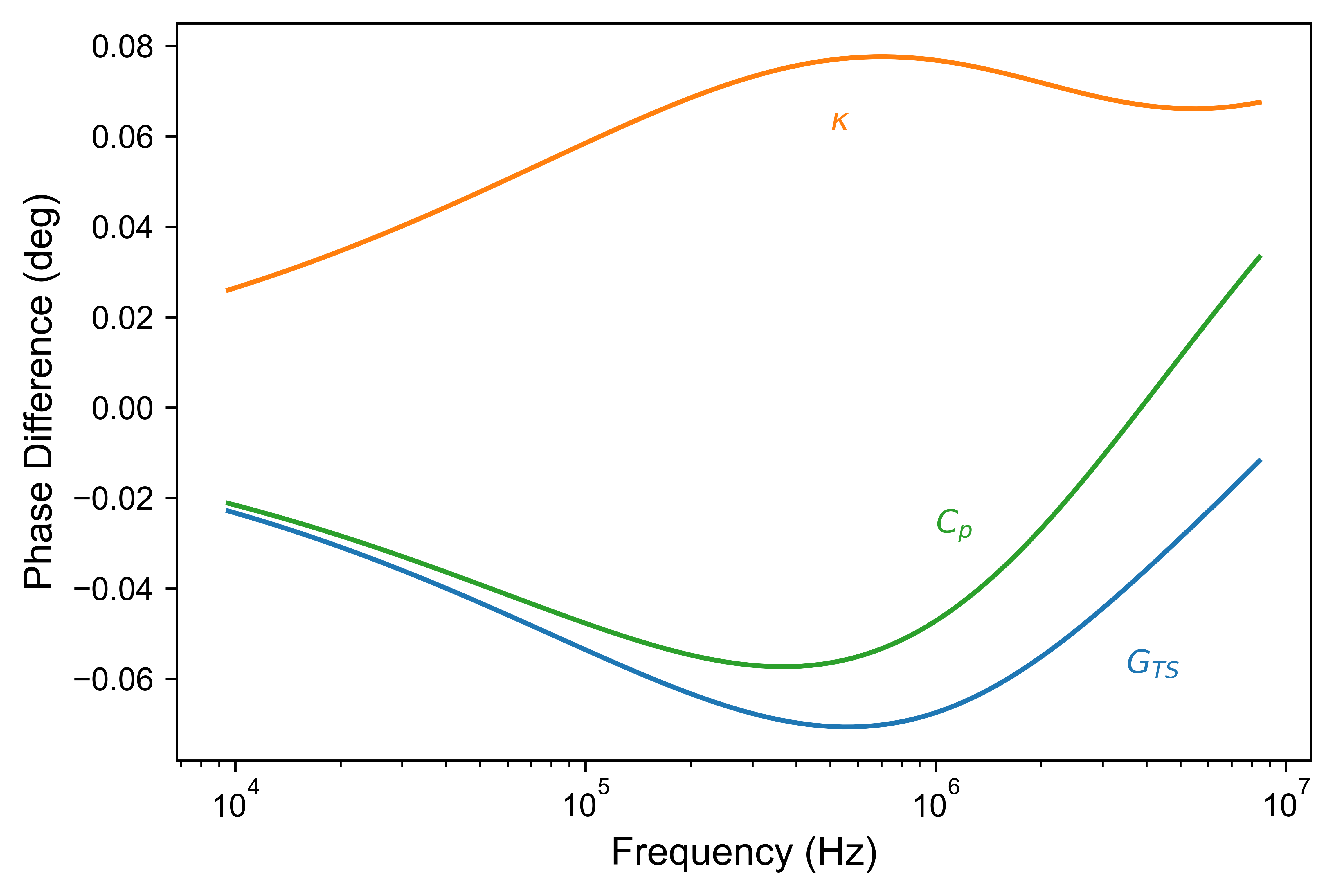}
    \caption{Sensitivity analysis of LaAlO$_3$ substrate parameters. These measurements were performed on a substrate without a film, coated in a transducer layer, and the results are shown in Fig.~3. Substrate thermal conductivity $\kappa$, volumetric heat capacity $C_p$, and the thermal boundary conductance between the transducer layer and the substrate $G_{TS}$ are shown.}
    \label{fig:lao_sens}
\end{figure}
\clearpage
\newpage
\section{Negligible Electron–Phonon Coupling Effects in the Substrate Measurements}

\begin{figure}
    \centering
    \includegraphics[width=\linewidth]{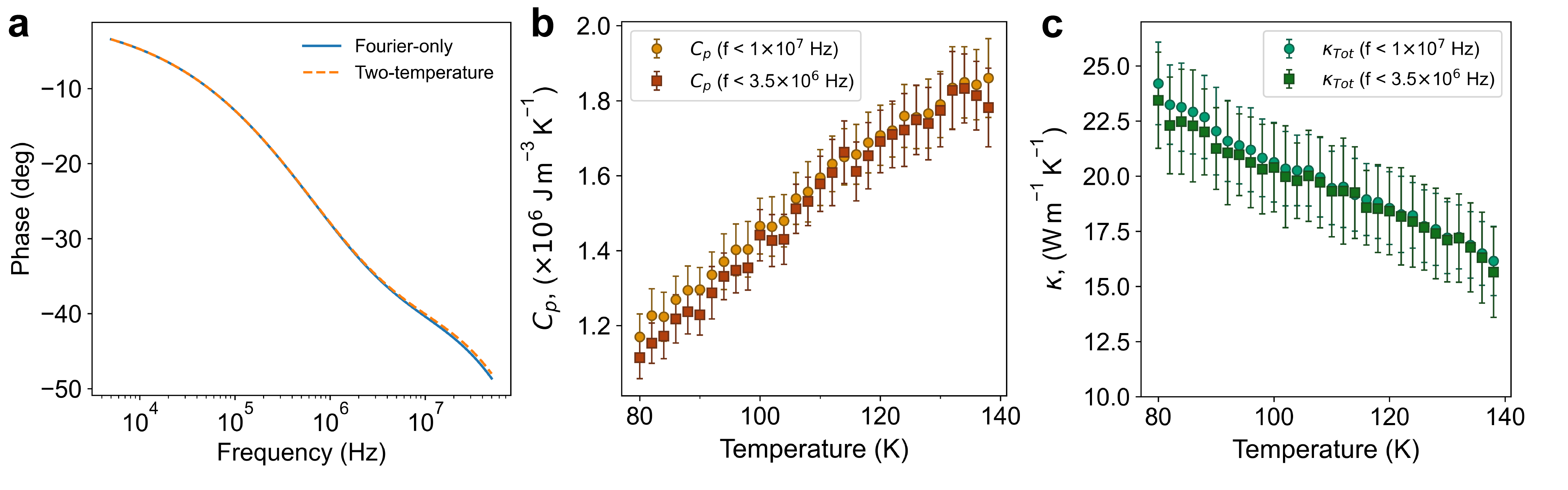}
    \caption{\textbf{a}, Modeled phase lag using the Fourier-only model (blue) and the two-temperature model including nonequilibrium electron-phonon coupling (orange) at 110 K. Parameters used: $\kappa_{\text{Au},e} =   112.8$ W m$^{-1}$K$^{-1}$, $\kappa_{\text{Au},p}= 4$ W m$^{-1}$K$^{-1}$, $C_{p,\text{Au}}=2.16\times 10^6$ J m$^{-3}$K$^{-1}$, $g=2\times 10^{16}$ W m$^{-3}$K$^{-1}$, $\kappa_{\text{LaAlO3},e} = 0$ W m$^{-1}$K$^{-1}$, $\kappa_{\text{LaAlO3},p} =19.46$ W m$^{-1}$K$^{-1}$, $C_{\text{LaAlO3},p}=1.60\times 10^6$ J m$^{-3}$K$^{-1}$, $G_{pp}=1.7\times 10^8$ $\mathrm{W m^{-2}K^{-1}}$. 
    \textbf{d}, Best fits for heat capacity, and \textbf{e}, thermal conductivity with using data with modulation frequencies $f<1\times10^7$ Hz (the fits used in the main text), and with $f<3.5\times10^6$ Hz. 
   We find that the best fits obtained using data with $f<1\times10^7$ Hz are similar to those using $f<1\times10^7$, a cutoff that  almost completely eliminates the influence of electron-phonon coupling.}
   \label{fig:elph}
\end{figure}

\begin{figure}[h]
    \centering
    \includegraphics[width=0.7\linewidth]{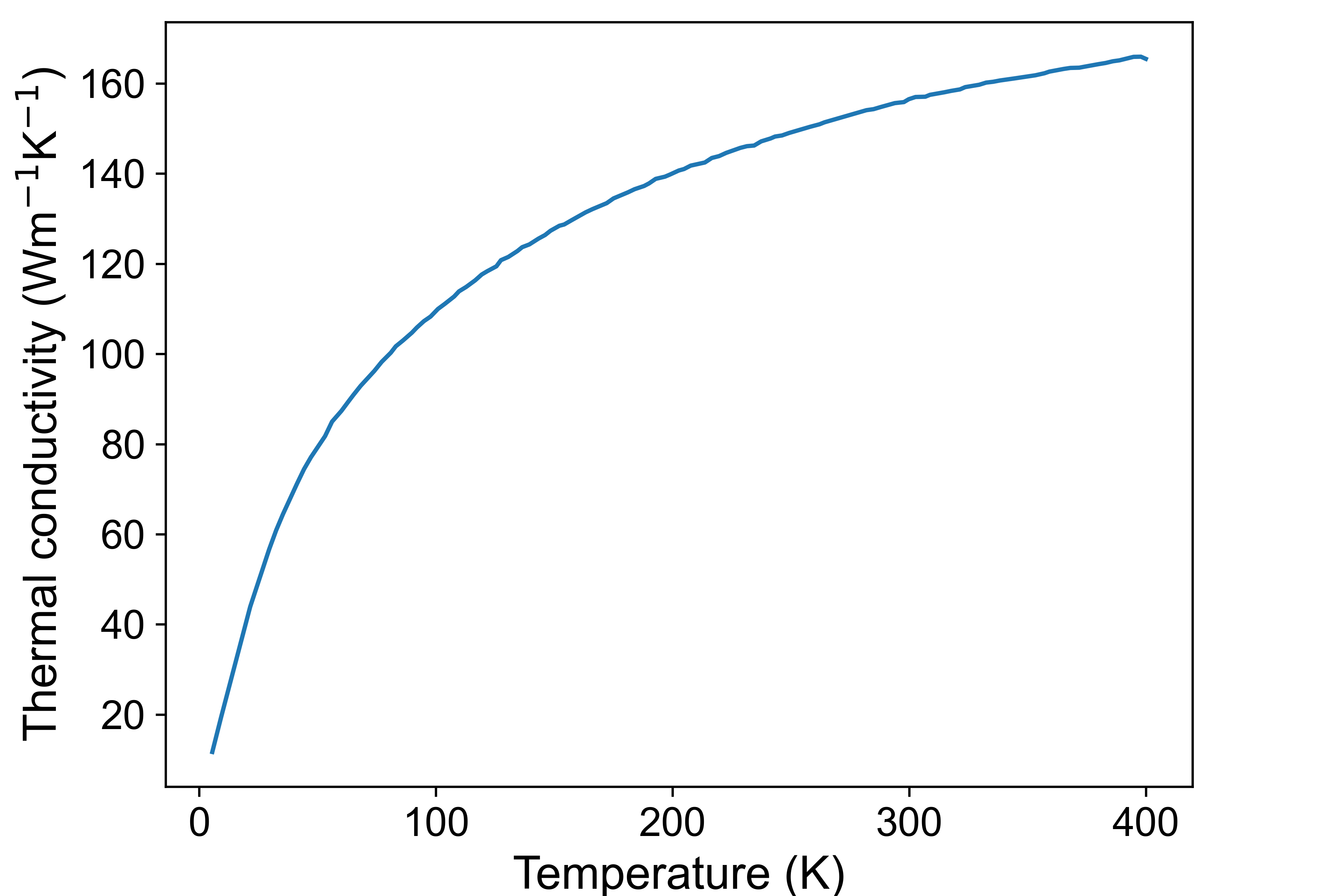}
    \caption{Thermal conductivity of 100 nm Au on glass using the Wiedemann-Franz law. The electrical conductivity of the gold transducer layer is measured with a four-point probe geometry using a Quantum Design Physical Property Measurement System.}
    \label{fig:goldkappa}
\end{figure}

We quantify the effect of nonequilibrium electron-phonon coupling in the gold transducer layer on the phase lag using a two-temperature model\cite{code,wilson2013} to determine a modulation frequency regime where the Fourier-only model is sufficient. 
We observe that the effects of electron-phonon coupling are more pronounced as the modulation frequency becomes higher. The fits in Fig.~3a use a maximum frequency of $10^7$ Hz. We reproduce those fits in Figs.~\ref{fig:elph}b,c as well as fits on the same data using a cutoff frequency of $3.5\times10^6$ Hz, and we do not observe a significant difference between them. We therefore conclude that a Fourier-only model with $f<10^7$ Hz is sufficient to exclude the influence of noneuilibrium electron-phonon coupling.

\clearpage

\section{Thermal Conductivity of the Gold Thin Film}

\clearpage
\section{Heat Capacity of NdNiO$_3$}
\begin{figure}[h!]
    \centering
    \includegraphics[width=0.7\linewidth]{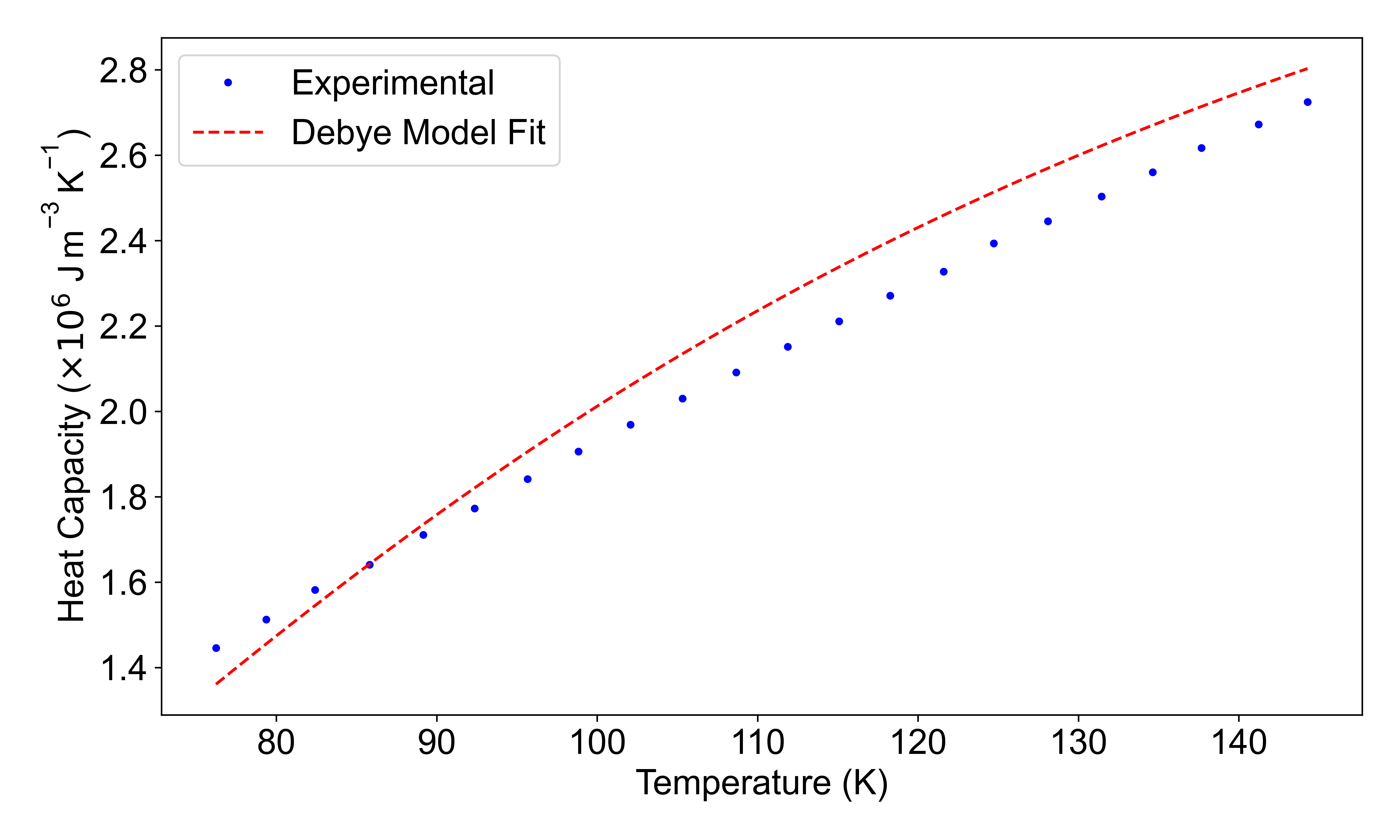}
    \caption{Heat capacity of NdNiO$_3$ as used to fit both the FDTR and FDPR  data. Blue dots represent volumetric heat capacity taken from literature measurements on bulk NdNiO$_3$, obtained using semi-adiabatic calorimetry on a PPMS\cite{hooda_electronic_2016}. Although the phase transition occurs at a higher temperature in the bulk NdNiO$_3$ and the epitaxy strain in thin-film NdNiO$_3$ can alter the lattice dynamics, we believe that these factors only minimally affect the sensible component (i.e., the baseline)  of the heat capacity. The red dashed line represents the Debye heat capacity prediction used to fix the heat capacity parameter in FDTR and FDPR measurements.}
    \label{fig:si_nickelate_cp}
\end{figure}

\newpage
\section{Data for Fig.~3b: Thermal Conductivity vs Temperature}

The following tables provide the measured thermal conductivity $\kappa$ as a function of temperature $T$, along with the associated uncertainty $\Delta\kappa$, for both cooling and warming cycles as shown in Fig.~3b. The data were averaged over three cooling cycles and two warming cycles (see Fig.~\ref{fig:kappa_unaveraged}).

\begin{table}[h!]
\centering
\caption{Thermal conductivity during cooling and warming}
\vspace{0.5em}
\begin{minipage}{0.48\textwidth}
\centering
\textbf{(a) Cooling}
\vspace{0.5em}

\begin{tabular}{ccc}
\hline
$T$ (K) & $\kappa$ (W\,m$^{-1}$K$^{-1}$) & $\Delta\kappa$ \\
\hline
83.30  & 1.87 & 0.135 \\
92.67  & 2.35 &  0.183 \\
98.68  & 2.34 & 0.222 \\
103.72 & 2.55 & 0.166 \\
108.00 & 2.26 & 0.275 \\
110.00 & 2.31 & 0.332 \\
113.05 & 2.62 & 0.197 \\
115.00 & 3.04 & 0.423 \\
118.00 & 3.20 & 0.548 \\
123.30 & 3.29 & 0.332 \\
133.32 & 3.71 & 0.320 \\
\hline
\end{tabular}
\end{minipage}
\hfill
\begin{minipage}{0.48\textwidth}
\centering
\textbf{(b) Warming}
\vspace{0.5em}

\begin{tabular}{ccc}
\hline
$T$ (K) & $\kappa$ (W\,m$^{-1}$K$^{-1}$) & $\Delta\kappa$ \\
\hline
83.30 & 1.87 & 0.146 \\
93.55 & 2.02 & 0.222 \\
97.93& 2.23 & 0.270\\
102.91 & 2.39 & 0.195 \\
113.15 & 2.82 & 0.351 \\
114.95 & 2.62 & 0.221\\
117.15 & 2.99 & 0.379 \\
119.10 & 3.17  & 0.354 \\
127.38 & 3.33 & 0.405 \\
131.17 & 3.52 & 0.485\\
133.14 & 3.55 & 0.331 \\
137.39 & 3.52 & 0.375 \\
\hline
\end{tabular}
\end{minipage}
\end{table}

\bibliography{references}